%% file: ms.tex
\documentclass[onecolumn, 11pt]{article}

\pdfoutput=1

\input{preamble.tex}

\begin{document}

\singlespace

\title{Network Effects and Incumbent Response to Entry Threats:\\Empirical Evidence from the Airline Industry}

\author{\textsc{Steve Lawford}\footnote{ENAC (University of Toulouse), 7 avenue Edouard Belin, CS 54005, 31055, Toulouse, Cedex 4, France (email: steve.lawford@enac.fr). I thank participants at the Conference on Complex Systems (CCS2023, Salvador, Bahia, Brazil), the 27th ATRS World Conference (2024, Lisbon, Portugal), and the 2nd International Conference on Empirical Economics (2024, Penn State, USA), for helpful suggestions.}
}

\date{}

\maketitle

\vspace{0.5cm}

\renewenvironment{abstract}{
  \noindent\textit{Abstract}---\ignorespaces
}{\par\noindent}

\begin{abstract}
    \noindent I investigate how incumbents in the U.S. airline industry respond to threatened and actual route entry by Southwest Airlines. I use a two-way fixed effects and event study approach, and the latest available data from 1999--2022, to identify a firm's price and quantity response. I find evidence that incumbents cut fares preemptively (post-entry) by 6--8\% (16--18\%) although the significance, pattern, and timing of the preemptive cuts are quite different to Goolsbee and Syverson's (2008)\nocite{goolsbee_syverson08} earlier results. Incumbents increase capacity preemptively by 10--40\%, up to six quarters before the entry threat is established, and by 27--46\% post-entry. My results suggest a clear shift in firms' strategic response from price to quantity. I also investigate the impact of an incumbent's network structure on its preemptive and post-entry behaviour. While the results on price are unclear, a firm's post-entry capacity reaction depends strongly on its global network structure as well as the local importance (centrality) of the route.
    
\end{abstract}

\section{Introduction}
Economists have long been aware that incumbents may react preemptively to the threat of market entry by a competitor, in particular by cutting prices or increasing quantity to deter or accommodate entry, and that these effects can be substantial (for example, Porter, 1980\nocite{porter80}; Tirole, 1988\nocite{tirole88}). This phenomenon was first documented empirically for the U.S. airline industry by Goolsbee and Syverson (2008)\nocite{goolsbee_syverson08} who define the \emph{threat} of entry, as distinct from actual entry, in a way that can be taken to data. They consider a \emph{route} (market) as direct return passenger travel between two airports A and B, the potential \emph{entrant} as Southwest Airlines, and the \emph{incumbents} as other major airlines that already serve the route. Using data over the years 1993 to 2004, they noticed that the empirical probability of future entry by Southwest on a route increases dramatically when Southwest first starts to fly routes out of both A and B but, crucially, is not yet transporting passengers between A and B. They call this situation \emph{dual-presence}, and use the time $t_{0}$ at which it is first established to represent the start of the entry threat, distinct from Southwest's actual entry at time $t_{e}$. The threat to incumbents persists as long as Southwest maintains dual-presence. Goolsbee and Syverson argue that incumbent firms act preemptively by cutting average fares when faced with competition from Southwest, even before it enters the route, but that they do not use capacity changes as a strategic response. However, subsequent empirical studies of the U.S. airline industry using data up to 2014 have found conflicting evidence on the timing, magnitude and choice of strategic measures, even for the specific case of reaction to Southwest (for example, Daraban and Fournier, 2008\nocite{daraban_fournier08}; Goetz and Shapiro, 2012\nocite{goetz_shapiro12}; Gayle and Wu, 2013\nocite{gayle_wu13}; Tan, 2016\nocite{tan16}; Parise, 2018\nocite{parise18}; Ethiraj and Zhou, 2019\nocite{ethiraj_zhou19}; Kwoka and Batkeyev, 2019\nocite{kwoka_batkeyev19}; Ma, 2019\nocite{ma19}). Given this, and the numerous changes that have taken place in the airline industry over the last twenty years, I use the latest available data to empirically evaluate whether incumbents use price or quantity when faced by an entry threat or actual entry.

In this paper, I formulate a two-way fixed effects and event study approach using a new dataset of U.S domestic airline routes that were threatened by Southwest entry from 1999 to 2022. I find some evidence that incumbents still cut fares preemptively, but the estimated effects are smaller, and the pattern and timing of the preemptive effects are quite different, to those reported by Goolsbee and Syverson\nocite{goolsbee_syverson08}, and the significance of the point estimates is not robust to the choice of standard errors (robust or clustered). However, there is strong evidence for post-entry fare cutting in all specifications. In my baseline model, the point estimates show significant fare reductions of up to 8\% in the two quarters \emph{after} the threat first arises at time $t_{0}$, but before Southwest enters the route at $t_{e}$, with a total post-entry fare cut of 18\%. Nearly half of the total reduction in average fares takes place before Southwest enters the route. The size of this relative effect agrees closely with Goolsbee and Syverson\nocite{goolsbee_syverson08}, although they report a larger preemptive fare drop of 17\% at $t_{0}$, and a total post-entry fare drop of 29\%. Furthermore, they show that significant preemptive price cutting starts as much as one or two years \emph{before} the threat first occurs at $t_{0}$. They conclude that incumbents will be aware of Southwest's intention to enter an airport well before commencement of actual operations because of public announcements, advertising, recruitment, advance sale of tickets, and other airport-level negotiations that are observable to an industry insider, and so they will \emph{anticipate} the start of the threat. I find no empirical support for this conclusion in my data. Instead, I argue that incumbents do use price to react preemptively to competition by Southwest but that they do not cut fares in anticipation of the start of the threat, waiting instead until dual-presence has been established. Goolsbee and Syverson\nocite{goolsbee_syverson08} do not find a significant quantity reaction to Southwest, although their point estimates indicate that incumbents increase the number of available seats on a threatened route. However, I observe highly significant preemptive and post-entry increases in capacity in all specifications, starting six quarters \emph{before} the threat occurs. The number of available seats increases by 30\% on the start of the threat at $t_{0}$, and by 43\% on entry at $t_{e}$, relative to the excluded period two to three years before $t_{0}$.

\begin{figure}[t]\centering
    \includegraphics[width=.85\linewidth, keepaspectratio]{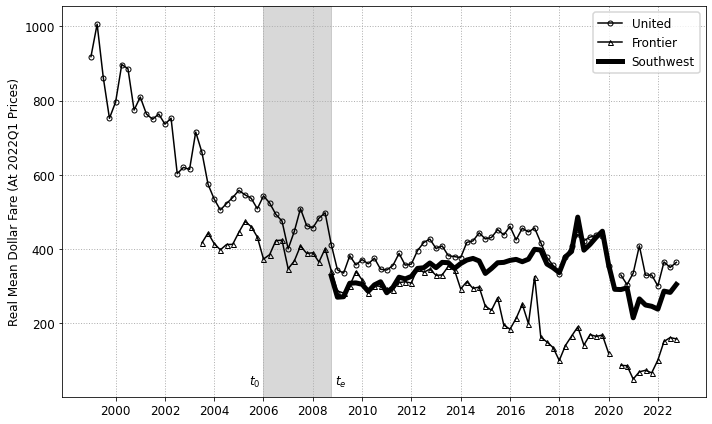}
    \caption{Real dollar fare. This figure depicts the mean real dollar fare (in 2022Q1 prices) for three carriers, United Airlines, Frontier Airlines and Southwest Airlines, on the route between Denver International Airport in Colorado and John Wayne Airport in California, between 1999 and 2022. The mean fare for each carrier-quarter is calculated across all direct return tickets in the data. Frontier established dual-presence and entered the route simultaneously in 2003Q3, and so it did not create a distinct threat for United. The shaded area represents the period during which Southwest had dual-presence on the route, starting at time $t_{0}$ and ending when Southwest actually entered the route at time $t_{e}$. Prior to $t_{0}$, Southwest was active at John Wayne but had no flights into or out of Denver. There is some evidence that United's and Frontier's fares fell during the period of Southwest's dual-presence. The activity of United and Frontier fell during the Covid-19 pandemic and neither carrier appears on this route in 2020Q2 after data pre-processing.}
    \label{fig:DEN_SNA_UA_F9_WN_fares}
\end{figure}

To fix ideas, it is useful to consider Fig. \ref{fig:DEN_SNA_UA_F9_WN_fares}, which plots the real mean dollar fare for travel between Denver International Airport in Colorado and John Wayne Airport in California, calculated across all direct return tickets in the data, for each carrier-quarter. The first incumbent, United Airlines, was present on the route from 1999 onward. Its real fares fell substantially over the next four years, although it did not face any threatened or actual competition on the route. In 2003Q3, Frontier Airlines established dual-presence and entered the route in the same quarter, and so it did not present United with a distinct entry threat. There is no obvious change in United's fares over the next couple of years, during which time it competed directly with Frontier for passengers. Southwest started operations at Denver in 2006Q1. Before then it was active at John Wayne and so this marked the establishment of dual-presence on the route, at time $t_{0}$. Southwest then entered the route with actual service in 2008Q4, at time $t_{e}$. The shaded area represents the period of Southwest's dual-presence, which lasted for two-and-a-half years. United's and Frontier's fares fell during that time, and again after actual entry.

I use all threatened routes from 1999--2022 to identify how incumbent carriers react in general to (the threat of) competition by Southwest. I estimate preemptive and post-entry effects, up to two years before and after the start of dual-presence, and up to two years after entry. I include control variables in some model specifications to investigate whether fare cuts are robust to  route-level characteristics, in particular route-length and a temperature-based proxy for leisure versus business routes. In a second set of model specifications, I use graph-theoretic variables to investigate whether an incumbent's post-threat or post-entry reactions are influenced by its network structure. Intuitively, a firm's price or quantity reaction to potential or actual competition is a local decision that is taken at the route-level (for example, Goolsbee and Syverson, 2008\nocite{goolsbee_syverson08}, find that incumbents do not cut fares on neighbouring routes that are not directly threatened by Southwest but that serve the same metropolitan areas). I find that an incumbent's local \emph{and} global network structure can both play a role, preemptively and post-entry. While the results on price are unclear, I observe greater preemptive fare cuts (in some quarters) on routes that are locally less ``central'', and for carriers with more ``spread out'' or ``clustered'' route networks. There is stronger evidence for the influence of network structure on a firm's quantity response. I find global and local post-entry effects, with greater capacity increases (measured by the number of available seats) for carriers with \emph{less} ``spread out'' or ``clustered'' route networks, and on routes that are locally \emph{more} ``central''. These network interaction variables are significant in all model specifications. A tentative conclusion is that incumbents (perhaps those with less efficient route networks) cut fares by more, preemptively, on routes that occupy a less central or less well-protected position in their network, while incumbents (perhaps those with more efficient route networks) increase capacity by more, post-entry, on routes that include a central (or hub) airport. In essence, the strongest strategic response is reserved post-entry for markets on which an incumbent is in the strongest position to respond.

\section{Literature}
In this paper, I make two distinct contributions. First, I use the latest available data on route networks and fares and available capacity, for passenger travel with U.S domestic airlines, to investigate the timing and pattern of incumbent price and quantity reactions to potential and actual route-level competition by Southwest. Second, I look for a relationship between an incumbent's strategic response and its global or local network structure. My work builds on two separate branches of the empirical literature: event studies, and specifically their application to entry threats (for example, Dafny, 2005\nocite{dafny05}; Ellison and Ellison, 2011\nocite{ellison_ellison11}; Tenn and Wendling, 2014\nocite{tenn_wendling14}; Cookson, 2018\nocite{cookson18}; Pan et al., 2019\nocite{pan_etal19}; Tomy, 2019\nocite{tomy19}; Wen and Zhu, 2019\nocite{wen_zhu19}; Wilson et al., 2021\nocite{wilson_etal21}); and the use of network-based measures as explanatory variables in econometric regressions, an approach that has been particularly popular in finance  (for example, Hochberg et al., 2007; 2010\nocite{hochberg_etal07}\nocite{hochberg_etal10}; Robinson and Stuart, 2007\nocite{robinson_stuart07}; Schilling and Phelps, 2007\nocite{schilling_phelps07}; Cattani et al., 2008\nocite{cattani_etal08}; Lindsey, 2008\nocite{lindsey08}; Ljungqvist et al., 2009\nocite{ljungqvist_etal09}; Phelps, 2010\nocite{phelps10}; Stuart and Kim, 2010\nocite{stuart_kim10}; Cai and Sevilir, 2012\nocite{cai_sevilir12}; Engelberg et al., 2012; 2013\nocite{engelberg_etal12}\nocite{engelberg_etal13}; Ozmel et al., 2013\nocite{ozmel_etal13}; Cohen-Cole et al., 2014\nocite{cohen-cole_etal14}; El-Khattib et al., 2015\nocite{el-khattib_etal15}).\\
\\
\noindent\emph{Comparison with Goolsbee and Syverson (2008)\nocite{goolsbee_syverson08}}\\
\\
\noindent I closely follow the methodology of Goolsbee and Syverson (2008)\nocite{goolsbee_syverson08}, who study the same basic ticket-level data that I use but for an earlier period 1993--2004, and I estimate a similar event study model. They observe 704 routes that were threatened by Southwest, 533 of which Southwest entered with direct service by the end of their sample. They restrict the incumbents to six large major carriers: American Airlines, Continental Airlines, Delta Air Lines, Northwest Airlines, Trans World Airlines (TWA) and US Airways, some of which no longer exist as distinct firms. They find that incumbents cut fares by 17\% when Southwest first establishes dual-presence at $t_{0}$, relative to a four-quarter window that ends two years before the start of dual-presence, and continue to cut fares for up to two years before Southwest actually enters, with a final 29\% fare cut in the three years following entry. They suggest that incumbents react in \emph{anticipation} of dual-presence, and report significant preemptive price cutting up to seven quarters \emph{before} time $t_{0}$. Their key argument for price as a strategic response is that incumbents use fares to try to lock-in existing customers, making it less likely that they will switch to Southwest when it finally enters the route. Their results are robust to including airport-specific operating costs in the regression, or changing the width of the event-window. They also show that fare cuts only occur on routes where there is a distinct threat of entry (for which $t_{0} \neq t_{e}$), and that there are no fare cuts on neighbouring routes that are not directly threatened by Southwest but that serve the same metropolitan areas. They find no evidence of a significant quantity reaction based on data on available seats (although their point estimates suggest a substantial increase in capacity) but report significant preemptive and post-entry increases in the load factor (the ratio of passengers to available seats).

I find evidence in the point estimates of preemptive and post-entry fare cuts that are roughly half as large as Goolsbee and Syverson\nocite{goolsbee_syverson08}, but no significant preemptive price reaction until the quarter \emph{after} dual-presence has been established using robust standard errors, and no significant preemptive price effect at all when errors are clustered by carrier-route. This suggests that incumbents do not anticipate the threat event, but wait at least one quarter after Southwest has started operating out of both endpoint airports, but is not yet flying the route. Compared to earlier results, one possible explanation is that incumbents are now in a stronger competitive position relative to Southwest, are able to wait longer before reacting preemptively, and then cut fares by less, although post-entry fare cuts remain substantial and highly significant in all specifications. However, when I look at capacity, the story is very different: I find significant preemptive capacity increases by incumbents up to six quarters before dual-presence, that continue post-entry, but no load factor effect. Together, my results point to a major change in the timing, magnitude and means of incumbents' strategic response over the last ten to fifteen years. There are two notable differences between Goolsbee and Syverson \nocite{goolsbee_syverson08} and my work that could contribute to these contrasting results, as well as more minor issues of dataset construction and the empirical filters that I use to find threatened routes.

First, our samples only overlap for the years 1999 to 2004. I find that there were very few threatened routes during that period and, in general, far fewer threatened routes over my full sample than Goolsbee and Syverson\nocite{goolsbee_syverson08} found over a shorter and earlier time frame. It is well-known that Southwest expanded its operations heavily in the 1990s, which suggests that dual-presence was first established then for many of the threatened routes in their sample.\footnote{Source: Southwest Airlines Press Releases swamedia.com/pages/1990-to-1995; and swamedia.com/pages/1996-to-2001 (accessed August 31, 2023).} However, there is a potential issue in the raw data during those years. From 1993 to 1998, Southwest reported each leg of its itineraries separately. One-way tickets were recorded in the same way as the two legs of direct nonstop return tickets, as well as any single leg of more complicated itineraries. I find no Southwest direct nonstop return tickets in the data during those years. It is not clear how to extract nonstop direct return tickets from 1993 to 1998, with their associated fares, so that they combine seamlessly with true nonstop direct return tickets from later data. If we assume that every itinerary leg is a one-way ticket and then double the fare (say) to obtain a return fare, that could add considerable noise to the data, even if we would expect the resulting route networks to be qualitatively similar. I avoid any ambiguity due to data-entry considerations by only using the years 1999 to 2022.

Second, there is a significant downward trend in the point estimates of the pre-threat indicators in Goolsbee and Syverson's (2008)\nocite{goolsbee_syverson08} baseline model for average logged fares, that includes controls for carrier-route and carrier-quarter fixed effects (see their Table II, column 1; and Table III). Fares fall by 13\% in the two years before $t_{0}$, relative to the earlier excluded period. Typically, a well-specified event study will have flat pre-treatment coefficients, which ensures that the treatment (here, the establishment of dual-presence at time $t_{0}$) was not anticipated, and it may be necessary to control explicitly for time trends in the dependent variable (see Miller, 2023\nocite{miller23}). Any pre-threat trend in real fares that is due to a general fall in the cost of air travel in the 1990s should be captured by Goolsbee and Syverson's\nocite{goolsbee_syverson08} carrier-quarter fixed effects. Fares did not fall equally across all routes either, relative to the start of dual-presence: there is no trend in pre-threat point estimates on ``nearby routes'' (see their Table IV, column 1) and there is inconclusive evidence of a downward trend on routes where $t_{0} = t_{e}$ so that the threat is not separate from entry (see their Table VI, column 2). If incumbents do indeed react years \emph{before} the establishment of dual-presence then it is harder to argue that the entry threat occurs precisely at $t_{0}$, since incumbents may well perceive a threat well before that time. In practice, this does not change the main message of Goolsbee and Syverson\nocite{goolsbee_syverson08} or my work, but while an incumbent may react preemptively (before entry) or post-entry, it is worth remembering that dual-presence is really the time at which an entry threat becomes more ``concrete''. Below, I make the distinction between ``pre-threat'' (before $t_{0}$) and ``post-threat'' (at or after $t_{0}$ but before $t_{e}$) preemptive reactions, as well as the unambiguous post-entry (at or after $t_{e}$) effects. Goolsbee and Syverson's\nocite{goolsbee_syverson08} results could also reflect an endogenous decision on the part of Southwest to establish dual-presence and then delay actual entry into routes on which the price of travel with the incumbents was already falling for other reasons but, given all the evidence, it is likely that incumbents do in fact anticipate the start of dual-presence in their sample. In my work, I find no significant pre-threat fare trends in the point estimates of the pre-threat indicators, in any model specification.\\
\\
\emph{Related papers on event studies}\\
\\
Tan (2016)\nocite{tan16} uses DB1B data to study the price response of six major incumbents to the entry of four low-cost carriers, including Southwest, for 1993--2009. He models logged mean one-way fares, with separate dummies for the quarter of entry and for each of the four subsequent post-entry quarters, including carrier-route and quarter fixed effects and route-quarter Herfindahl and carrier-quarter bankruptcy dummy control variables. (An entrant must not have served the route for 12 quarters before entry and it must remain on the route for at least eight quarters after entry.) Tan finds that incumbents cut fares by 10\% on Southwest's entry at $t_{e}$, and by up to 16\% post-entry at $t_{e}+3$, relative to an excluded period between 9 and 12 quarters before entry. Although Tan does not examine entry threats explicitly, his sample includes 1,286 routes entered by Southwest, which must include some of the threatened routes in Goolsbee and Syverson, and my work. Still, there is no evidence of a pre-entry trend or fare cutting in the four quarters before Southwest's entry: indeed, there is a small significant \emph{positive} price reaction of up to 2.5\% between two and four quarters before $t_{e}$. Tan reports qualitatively similar results on entry and post-entry dummies for the 10th and 90th fare percentiles (with larger price cutting for higher fares), and for other low-cost entrants.

Gayle and Wu (2013)\nocite{gayle_wu13} examine the price reaction of an incumbent carrier to an entry threat or actual competition, without any restrictions on the sets of carriers. They attempt to control for potential endogeneity between route-level errors and the number of carriers on the route by estimating a two-step cross-sectional structural model for profits and fares with directional round-trip DB1B data for the first and third quarters of 2007. They define a threat (dual-presence) as a carrier that serves both endpoints of a route in 2007Q1 but has not yet entered in 2007Q3. Gayle and Wu's\nocite{gayle_wu13} route-level baseline model for logged median fares (their Table 5, column 1) includes the \emph{number} of firms with dual-presence and the \emph{number} of actual firms serving the route in 2007Q3. They include controls for population, income, distance and airport slot restrictions. They report that incumbents cut fares by more when faced with actual competition than by an entry threat: logged median fares fall by 4.5\% when the number of carriers on a route increases by one, and by 1\% when incumbents face an additional threat. These effects are much smaller than the price reactions of major carriers to Southwest reported by Goolsbee and Syverson\nocite{goolsbee_syverson08}, Tan\nocite{tan16}, and in my work.  Gayle and Wu\nocite{gayle_wu13} find a larger price cut (2\%) following an additional threat when the potential entrant has a hub at one or both of the endpoints on the route, and a \emph{positive} price reaction (8\%) when the potential entrant is a member of the same airline alliance as one of the incumbents on the route (their Table 5, column 3). They find qualitatively similar results for the 25th and 75th fare percentiles, but there is evidence that the effect of actual competition is larger (6\%) for low fares and smaller (2\%) for high fares (on the other hand, the reaction to an entry threat is insignificant for low fares, and similar for high fares). Their results change very little after correcting for endogeneity.

Daraban and Fournier (2008)\nocite{daraban_fournier08} measure the fare reaction of seven major incumbents to route entry or exit by Southwest or other low-costs, where the fare is averaged across all the incumbents on a route. Their data includes one-way fares on unidirectional round-trips for 2,000 high-volume routes, from the DB1B for 1993--2006. Their baseline model has dummies for the four quarters before and after actual entry and exit, and the excluded period (route-quarters) is five quarters before entry on routes that experience entry, and routes where there was no entry. They report a 15\% drop in incumbent fares on Southwest's entry at $t_{e}$ (their Table 2, column 1), with a final post-entry drop of 20\% at $t_{e} + 3$ and $t_{e} + 4$, in line with my baseline results. While Daraban and Fournier do not directly model entry threats, they find a small but significant fare reduction of 4--6\% in the four quarters before entry, smaller than Goolsbee and Syverson but in line with my baseline results during dual-presence. Incumbent reactions before and after entry by a low-cost other than Southwest are more than 50\% smaller than those for Southwest.

Ethiraj and Zhou (2019)\nocite{ethiraj_zhou19} measure incumbent response to entry threat and actual entry by Southwest, for nonstop flights between the 100 highest-volume airports in 2003--2012, using data from the DB1B and the commercial OAG. Their baseline specification is similar to Goolsbee and Syverson's\nocite{goolsbee_syverson08}, with the addition of origin airport-quarter fixed effects, but their results differ in two important ways. First, they find no pre-entry price response for either full-service or low-cost incumbents (their Table 3, columns 1 and 2). Incumbents cut fares by 11--18\% post-entry. Second, there is a significant positive pre-threat, post-threat and post-entry trend in seat capacity for full-service incumbents: 30\% at $t_{0}$ and 42\% at $t_{e}$. (The capacity response is concentrated on routes with a large number of leisure passengers.) However, there is no capacity reaction by low-cost incumbents! (See their Table 3, columns 5 and 6.)

Kwoka and Batkeyev (2019)\nocite{kwoka_batkeyev19} use DB1B and T-100 data for 2011--2014 to examine the response of two incumbents, United Airlines and US Airways, to entry by several low-cost or major competitors. They focus on a small number of specific routes on which entry was exogenously determined by mandated slot or gate divestitures. Using a Differences-in-Differences approach, their baseline model includes indicators for five quarters before and after entry, and route and quarter fixed effects, with an excluded period six to nine quarters before entry. Unsurprisingly, they find substantial heterogeneity in incumbent response (fares, departures and available seats) across different routes, both before and after entry, including preemptive fare \emph{increases} when Southwest enters a \emph{neighbouring} route!

Goetz and Shapiro (2012)\nocite{goetz_shapiro12} explore flight codesharing as another possible strategic reaction to Southwest's entry, using a linear probability model and data on nonstop flight segments from the DB1B for 1998--2010. In their baseline model (their Table 2, column 3), which only includes dummies for the start of dual-presence and entry, the probability of an incumbent entering into a route-level codesharing agreement increases by 5--7\% when Southwest starts dual-presence at $t_{0}$ or enters at $t_{e}$. (Across their sample, the baseline probability of codesharing is 20\%.) In another specification, they include pre-threat, post-threat and post-entry dummies (their Table 7, column 1) and find significant increases in the codesharing probability of 5--8\% between $t_{0} - 1$ and $t_{0} + 2$, but the entry effect is no longer significant, and there are no post-entry effects. Parise (2018)\nocite{parise18} shows that a one standard deviation increase in entry threat by Southwest is associated with a 6 percentage point increase in the proportion of long-term debt (maturing in more than three years) over total debt held by incumbents, using T-100 and financial data for 1990--2014 (see his Table 5). His baseline model for airline-quarter debt maturity includes an airline-level entry threat that measures the fraction of an incumbent's market exposed to the likely entry of Southwest, and is calculated by aggregating city-pair route-level entry threats. Ma (2019)\nocite{ma19} finds that incumbents with lower financial leverage (debt-to-asset ratio) cut their prices more aggressively than those with higher leverage, when faced with an entry threat by Southwest. This is reversed when Southwest actually enters the route, at which point it is higher-leverage incumbents that cut fares more aggressively.\\
\\
\emph{Related papers on network measures}\\
\\
A number of papers in finance use local or global network measures as explanatory variables. Hochberg et al. (2007)\nocite{hochberg_etal07} examine the consequences of the network organizational structure of U.S. venture capital (VC) funds, where two funds are linked if they syndicate investment (co-invest) in the same portfolio company. They find that funds with more influential network positions, as measured by degree, closeness and betweenness centralities, and interactions between centrality and firm characteristics, have significantly better performance, and argue that better-quality relationships facilitate sharing of information and resources. Hochberg et al. (2010)\nocite{hochberg_etal10} study barriers-to-entry and show that more densely-networked (density) local VC markets experience less entry by outside VCs because of greater incumbent network externalities, and that incumbents who do facilitate entry by an outsider may be punished with reduced future network access. Ozmel et al. (2013)\nocite{ozmel_etal13} show that the performance of biotechnology companies that are backed by VC syndicates depends positively on the connections (Bonacich centrality) of the individual VC firms.

Strategic alliances facilitate information transmission and resource-sharing. Robinson and Stuart (2007)\nocite{robinson_stuart07} look at the network of past strategic alliances and joint ventures between biotechnology and pharmaceutical firms. They find that as two firms are more closely linked (the number of repeat edges over time, or the two-step paths between two nodes in a given period) or more central (Bonacich centrality), then equity participation falls and pledged funding from clients increases, through the impact of long-term reputational effects on specific contracts. Schilling and Phelps (2007)\nocite{schilling_phelps07} study strategic alliances between U.S. firms in high technology manufacturing industries. The number of successful patent applications increases when the alliance network is more clustered (transitivity) or less spread out (high reach, which is the mean across nodes of the sum of each node's inverse shortest path lengths to its neighbours). Phelps (2010)\nocite{phelps10} looks at strategic alliances between telecommunications equipment manufacturers and shows that the relative number of new patent citations increases in the technological diversity of a firm’s alliance partners, and that this effect is increasing in the firm's network density.

Social networks and personal ties matter too. El-Khattib et al. (2015)\nocite{el-khattib_etal15} argue that well-connected CEOs (degree, closeness, betweenness, eigenvector centralities) in S\&P 1500 firms enjoy better access to private information, and increased bargaining power and boardroom loyalty. They show that CEOs with higher network centrality pursue acquisitions more often, and that these merger deals are more likely to destroy shareholder value, and contribute to managerial entrenchment, with high-centrality CEOs less likely to be dismissed if a merger generates losses. Cai and Sevilir (2012)\nocite{cai_sevilir12} find that acquirer firms in M\&A transactions obtain higher annoncement returns when there is a current board connection between the target and acquirer (at least one common director, or at least one director from the target and acquirer firms are members of the board of a third firm). Engelberg et al. (2013)\nocite{engelberg_etal13} find that the number of a CEO's personal connections to high-ranking executives or directors at other firms is a strong predictor of their salary and total compensation. CEOs with large social networks earn more than those with smaller networks. (They define a CEO's ``rolodex'', or degree, as the sum of old professional, university and social connections.) Engelberg et al. (2012)\nocite{engelberg_etal12} study U.S. commercial banks and public firms and provide evidence that personal relationships (at least one school connection or third-party past professional tie between the borrower and any syndicate bank) lead to more favourable financing terms, in particular, reduced interest rates, fewer covenants, and larger loan amounts. Stuart and Kim (2010)\nocite{stuart_kim10} find that U.S. public firms are more likely to be targeted in a private-equity-backed transaction when they have more directors who have served as directors or executives of other public firms, or who have served on the board of a company that has previously attracted a take-private offer (degree).

\newpage
\section{The Data}
\noindent\emph{Route network data and real fares}\\
\\
\noindent My main data source is the DB1B, a publicly-available collection of airline ``tickets'' (itineraries) for passenger travel on the U.S. domestic network, over the years 1999 to 2022. It contains quarterly demand-side information on operating and ticketing carrier, origin and destination airport, type of ticket, connections, and nominal dollar fares. I combine the DB1B with monthly supply-side data on flight segments from the T-100, a public database that includes available capacity (number of seats offered). I downloaded the DB1B and T-100 from the U.S. Department of Transportation's TranStats website in August 2023, and then implemented filtering and aggregation pre-processing algorithms to build carrier route networks. I give a full description of the dataset construction in Appendix \ref{sec:dataprocessing}.

I keep direct nonstop round-trip tickets for scheduled coach class (economy) travel between U.S. airports with a U.S. carrier. I drop very low and very high fares, as well as carrier-routes with a small number of passengers and carriers with very small networks. A typical observation in my route network data is a nonstop unidirectional round-trip from Washington Dulles (IAD) to Los Angeles (LAX), operated by United Airlines in 2022Q1, with 2,591 passengers recorded in the DB1B (this is much lower than the actual number of passengers because the DB1B is a 10\% random sample of all tickets). The nominal fares range from \$20 (the low-fare cut-off) to \$1,697, with a median fare of \$381. The load factor is 76\%, computed as the ratio of total number of passengers to available seats, both recorded in the T-100, across all ticket types (in contrast to the DB1B, the T-100 is a 100\% census of flight segments). I repeat these steps for each time period to give quarterly carrier route networks, augmented with route-level characteristics including the real fare distributions and the number of passengers and available seats. In much of the literature, it is unclear whether nominal or real fares are being modelled. I transform nominal fares to real fares at 2022Q1 levels using price index data from the U.S. Bureau of Labor Statistics.\\
\\
\noindent\emph{Routes threatened by Southwest entry}\\
\\
\noindent I follow Goolsbee and Syverson (2008)\nocite{goolsbee_syverson08} and define \emph{no-presence} as Southwest not being present in either of the endpoint airports on a route, \emph{single-presence} as Southwest operating out of one of the endpoint airports on the route but not both, \emph{dual-presence} as Southwest operating out of both of the endpoint airports but not flying the route itself, and \emph{entry} as the start of Southwest's actual service on the route. I assign one of these categories to each route-quarter in the data. In my work, I keep a \emph{threatened route} if it satisfies all of the following conditions:

\begin{enumerate}

\item At some point over the full sample, Southwest is in both single-presence (starting at time $t_{s}$) and dual-presence (starting at time $t_{0}$), and then enters the route (at time $t_{e}$). I require $t_{s} < t_{0} < t_{e}$ and I only keep the first series of these events on a given route in case dual-presence (or entry) occur repeatedly.

\item Southwest's presence on a route evolves monotonically, so that it is continually in single-presence from $t_{s}$ to $t_{0}-1$, and then continually in dual-presence from $t_{0}$ to $t_{e}-1$, and then serves the route continually from $t_{e}$.

\item At least one incumbent offers passenger travel on the route in each time period across the twenty-five quarter window $[ t_{0} - 12, t_{0} + 12 ]$ around the start of dual-presence, and also at $t_{e}$. I also keep all contiguous time periods from $t_{e} + 1$ up to $t_{e} + 12$ that are served by the incumbent, truncating the post-entry window as soon as the incumbent no longer serves the route (even if it returns to the route later), or at the end of the sample.

\end{enumerate}

\begin{figure}[t]
        \centering
        \includegraphics[width=.85\linewidth, keepaspectratio]{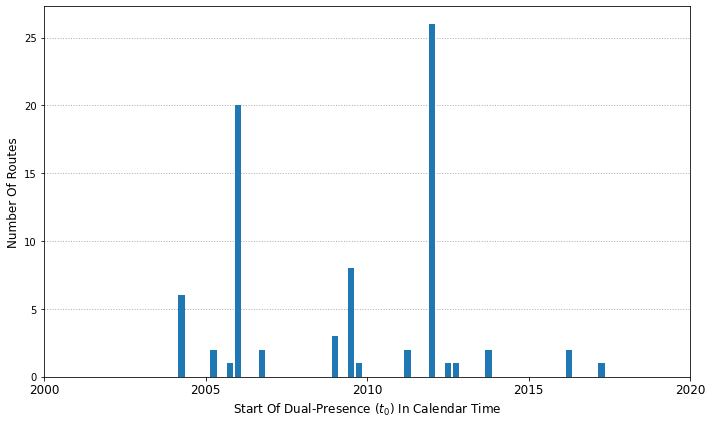}
        \caption{Threat event time. Southwest threatens 78 routes across the full sample. This figure shows the quarter $t_{0}$ at which the threat (as measured by dual-presence) is first established, in calendar time. I keep all routes on which Southwest establishes a threat during the sample, and later enters at time $t_{e}$, and on which at least one incumbent has continual service in a three-year window on either side of $t_{0}$, and also at $t_{e}$. During the Covid-19 pandemic, some carriers appear to have no service after data pre-processing, and so I find no threatened routes after 2019.}
        \label{fig:WN_dual_presence_date_hist}
\end{figure}

\begin{figure}\centering
    	\includegraphics[width=.85\linewidth, keepaspectratio]{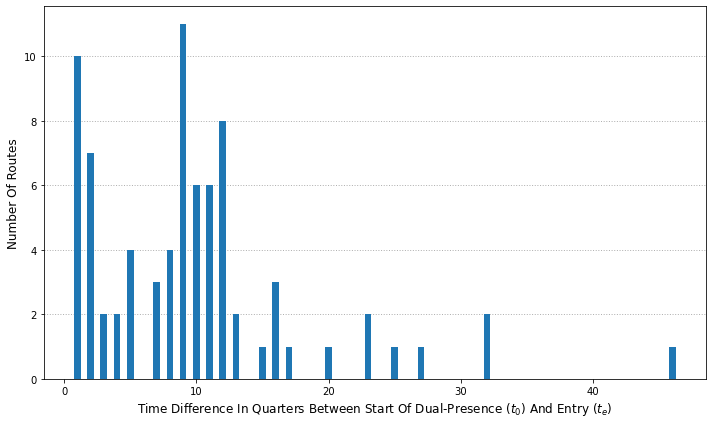}
    	\caption{Time from threat to entry. This figure shows the time difference $t_{e} - t_{0}$, in quarters, between the start of dual-presence by Southwest and its actual entry on a route. Southwest enters within three years on 56 out of 78 threatened routes.}
    	\label{fig:WN_entry_dual_presence_difference_hist}
\end{figure}

\noindent With these conditions, threat and entry events are observed separately, and there is no ambiguity in the progression from one state to the next. I am able to observe any pre-threat incumbent reactions that anticipate the start of dual-presence, and the consequences of remaining in the states of dual-presence or entry. From 1999 to 2019, there are 3,232 routes and 182,702 route-carrier-quarters in my data. Southwest entered 37 airports and it exited 9 airports across the sample: San Francisco (SFO) in 2001, George Bush  (IAH) in 2005, Branson (BKG) and Key West (EYW) and Jackson-Medgar Wiley Evers (JAN) in 2014, Greenville-Spartanburg (GSP) in 2016, Akron-Canton (CAK) and Dayton (DAY) in 2017, and Bishop (FNT) in 2018.\footnote{Goolsbee and Syverson (2008, p.1614, footnote 5)\nocite{goolsbee_syverson08} also note Southwest's departure from SFO in 2001.} Of these, it re-entered SFO in 2007 and GSP in 2019. Southwest threatens entry into 78 routes in the data, served by 11 different incumbents: AirTran Airways (FL), Alaska Airlines (AS), Allegiant Air (G4), American Airlines (AA), Delta Air Lines (DL),  Frontier Airlines (F9), JetBlue Airways (B6), Spirit Airlines (NK), Sun Country Airlines (SY), United Airlines (UA), and US Airways (US). The data forms an unbalanced panel with 98 threatened carrier-routes and 87 quarters. In total, there are 3,157 threatened route-carrier-quarter observations. Fig. \ref{fig:WN_dual_presence_date_hist} shows the distribution of threat event times $t_{0}$, which are spread over the years 2004 to 2017. Nearly 60\% of the routes that are threatened follow from Southwest's entry into just two airports: Denver (DEN) in 2006 and Hartsfield-Jackson Atlanta (ATL) in 2012. Fig. \ref{fig:WN_entry_dual_presence_difference_hist} presents the distribution of the time between the start of dual-presence and entry, in quarters. It shows that Southwest entered more than 70\% of threatened routes within three years of the start of dual-presence.

\section{Empirical Analysis}
I estimate the following reduced form log-linear regression:
\begin{equation}\label{eq:baseline}
\ln \, y_{\, ij \, t} = \left(\sum_{k \in \{-8, \ldots, -1, 0, 1, 2, 3+\}}\gamma_{\, k} \cdot D_{t_{0}(j) + k}\right) + \left(\sum_{k \in \{0, 1+, 3+\}}\gamma_{\, k} \cdot D_{t_{e}(j) + k}\right) + \alpha_{\, ij} + \delta_{\, t} + \beta \cdot X_{\, ij \, t} + \epsilon_{\, ij \, t}.
\end{equation}
The $y$ variable is the incumbent's outcome on a threatened route. In my baseline model, this is the mean \emph{real} fare across all direct return tickets, given incumbent carrier $i$, route $j$ and quarter $t$. In other specifications, it is a real fare percentile, the total number of passengers or available seats, or the load factor. The preemptive indicator variables $D_{t_{0}(j) + k}$ capture each of the 9 quarters up to and including the start of dual-presence at $t_{0}$, and each of the 2 quarters after $t_{0}$, and there is a single indicator for between 3 and 12 quarters after $t_{0}$ (the latter is denoted $k = 3+$). The post-entry indicator variables $D_{t_{e}(j) + k}$ capture entry at $t_{e}$, with a single indicator for between 1 and 2 quarters after $t_{e}$ (denoted $k = 1+$), and a single indicator for between 3 and 12 quarters after $t_{e}$ (denoted $k = 3+$). The preemptive and post-entry coefficients $\gamma_{\, k}$ measure the variation in the incumbent's outcome over time, relative to an excluded benchmark period between 9 and 12 quarters before $t_{0}$. The indicators are mutually exclusive and so these effects are not additive, and the approximate percentage change in $y$ is given by $100(\exp(\gamma_{\, k})-1)$.

\begin{figure}[t]
        \centering
        \includegraphics[width=.75\linewidth, keepaspectratio]{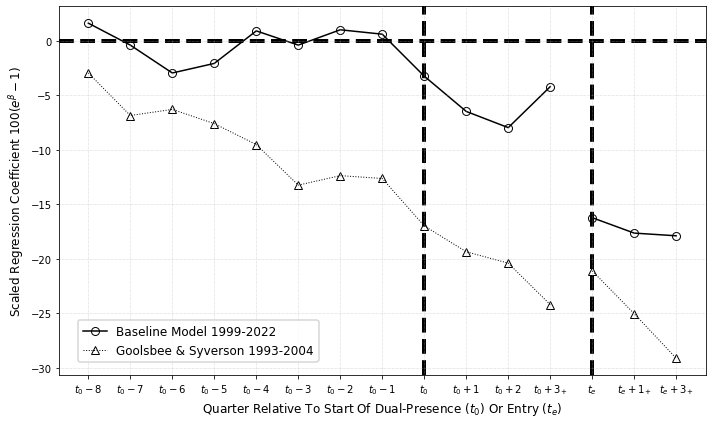}
        \caption{Percentage change in mean real fare relative to threat and entry events. This figure shows the standardized point estimates $100(e^{\beta}-1)$ of quarterly dummies relative to the start of Southwest's dual-presence at $t_{0}$ and actual entry at $t_{e}$. The solid line gives estimated percentage changes from my baseline model (\ref{eq:baseline}) that are computed from the point estimates $\beta$ reported in the first column of Table \ref{tab:baseline_model}. For comparison, the dashed line gives estimated percentage changes from the first column of Goolsbee and Syverson's (2008, Table II) baseline model.}
        \label{fig:WN_regression_coefficients_model_1}
\end{figure}

Dual-presence and entry can occur at different calendar times on different routes, and so the threat and entry event times $t_{0} = t_{0}(j)$ and $t_{e} = t_{e}(j)$ are functions of the route. For a given carrier-route, the time $t = t(i,j)$ is in the set $\{t_{0} - 12, \ldots, t_{0}, \ldots \min(t_{e}, t_{0} + 12)\} \cup \{t_{e}, \min(t_{e} + \tau, t_{e} + 12)\}$, where $t_{e} + \tau$ is the end of the full sample. I include carrier-route fixed effects $\alpha_{\, ij}$ and quarter fixed effects $\delta_{\, t}$ in my baseline model, which should capture any time trends in real fares. Goolsbee and Syverson (2008)\nocite{goolsbee_syverson08} include carrier-quarter fixed effects and not quarter fixed effects. I am unable to do this because of multicollinearity. They also cluster standard errors at the carrier-route level. In my baseline model I use standard errors clustered by carrier-route. I report results with robust standard errors in Appendix \ref{sec:appendix_robustness}. There is growing evidence that robust standard errors can be too small while clustered standard errors can be too large, leading to unnecessarily wide clustering confidence intervals in some common situations, even when there is an apparent pattern in the point estimates (for example, Abadie et al., 2023\nocite{abadie_etal23}).\footnote{In one case, Goolsbee and Syverson (2008, p.1621) argue for a pattern in logged passengers around $t_{0}$ even though all of the point estimates are insignificant: ``The estimates are imprecise, but the point estimates suggest that passenger traffic rises on threatened routes in the period before and around when Southwest enters the second endpoint airport.'' (See their Table II, column 2.) My results support this.} In the discussion I make it clear when I am drawing a conclusion based on clustered significance, or on the basis of point estimates alone (when they are still significant by robust standard errors but not by clustered standard errors). In some specifications, I include control variables $X_{\, ij \, t}$, for route-length or a temperature-based proxy for leisure versus business routes. In the estimation of equation (\ref{eq:baseline}), I weight observations by the number of passengers recorded in the DB1B for each carrier-route-quarter.\\
\\
\noindent \emph{Evidence for a price reaction}\\
\\
I report the results for logged mean real fares and percentiles in Table \ref{tab:baseline_model_clustered_ses} (clustered) and Table \ref{tab:baseline_model} (robust). In column 1, there is no evidence for a significant pre-threat trend that anticipates dual-presence, nor in the point estimates themselves. The point estimates suggest a moderate post-threat effect, whereby incumbents cut mean fares in a ``U-shaped'' pattern as long as dual-presence is maintained, by 6--8\% in the two quarters after $t_{0}$, which for a fare of \$400 is between \$26 and \$32. If dual-presence continues beyond two quarters without Southwest entering the route, mean fares fall by 4\% relative to the excluded period, suggesting that the price reaction to the threat diminishes the longer it takes for Southwest to enter. The post-threat reaction is not significant under clustering. When actual entry takes place at $t_{e}$, mean fares fall by 16\%, with a final post-entry reduction of 18\%, which for a fare of \$400 is around \$72. The post-entry coefficients are economically-important and highly significant in all specifications. Incumbents react convincingly when faced with actual competition, as expected, but there is weaker evidence of a preemptive reaction which suggests that firms do not make strong use of price to deter or accommodate Southwest's entry by reinforcing customer loyalty in an effort to reduce post-entry demand for the entrant. Fig. \ref{fig:WN_regression_coefficients_model_1} compares the estimated percentage changes in average real fares relative to threat and entry events at $t_{0}$ and $t_{e}$ from Table \ref{tab:baseline_model} to Goolsbee and Syverson's (2008, Table II)\nocite{goolsbee_syverson08} baseline model. It shows the absence of a pre-threat trend in my baseline model and smaller post-threat and post-entry effects compared to Goolsbee and Syverson\nocite{goolsbee_syverson08}.

I also investigate the incumbent response across the fare distribution, for the 10th, 25th, 75th and 90th percentiles. In columns 2 to 4 of Table \ref{tab:baseline_model}, there is very weak evidence of fare cutting (up to 5\%) before dual-presence, but for isolated quarters, there is no consistent pattern in the point estimates, and the significance vanishes under clustering. In column 5 of Table \ref{tab:baseline_model}, the positive sign of a pre-threat effect at $t_{0}-2$ is unexpected: it is not clear why incumbents would increase fares at that time, and there are no other pre-threat movements. I conclude that there is no good evidence for a pre-threat reaction across the fare distribution. The point estimates again suggest a ``U-shaped'' pattern of post-threat fare cutting that are largest around $t_{0}+2$: low fares fall by 6\% (columns 2 and 3) while high fares fall by 8--11\% (columns 4 and 5). The high fare reactions are much more significant than the low fare reactions, and are in several cases robust to clustering. As for mean fares, there are highly significant post-entry fare cuts across the fare distribution of between 14--19\%; the biggest reductions occur at $t_{e}$ for low fares but later (between $t_{e}+3$ and $t_{e}+12$) for high fares. These results suggest that post-threat fare cutting may be more pronounced and persistent for high fares, while there are clearly significant post-entry fare reductions across the fare distribution.

In column 6 of Table \ref{tab:baseline_model}, I show that the results from my baseline model for logged mean fares are robust to distance and temperature controls. I include route-level distance (in hundreds of miles) and distance-squared. As a proxy for leisure routes, I also control for the annual absolute temperature differential between the average January temperature, in Fahrenheit, of the endpoint states of the route. I expect the absolute temperature differential to increase in the percentage of leisure (rather than business) passengers on a route. Both sets of controls are highly significant and have the expected signs, in all specifications. Holding everything else constant, mean fares increase in distance to 890 miles, after which they fall; and a ten-degree Fahrenheit increase (or a 5.5\% degree Celcius increase) in temperature differential reduces average fares by 3\%. With controls, the point estimates of all preemptive and post-entry indicators have the same signs as the baseline model, and all post-threat effects are about 1\% lower than the baseline, and are highly significant in all specifications.

\begin{figure}[t]
        \centering
        \includegraphics[width=.75\linewidth, keepaspectratio]{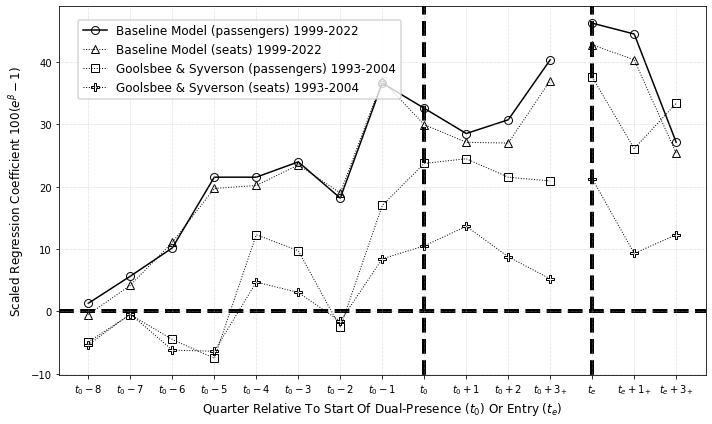}
        \caption{Percentage change in passengers (T-100) or available seats (T-100) relative to threat and entry events. This figure shows the standardized point estimates $100(e^{\beta}-1)$ of quarterly dummies relative to the start of Southwest's dual-presence at $t_{0}$ and actual entry at $t_{e}$. The estimated percentage changes from my baseline model (\ref{eq:baseline}) that are computed from the point estimates $\beta$ reported in columns 1--2 of Table \ref{tab:baseline_model_T100_robust}. For comparison, I include estimated percentage changes from columns 1--2 of Goolsbee and Syverson's (2008, Table V) baseline model.}
        \label{fig:WN_regression_coefficients_model_1_passengers_seats}
\end{figure}

\begin{figure}[t]
        \centering
        \includegraphics[width=.75\linewidth, keepaspectratio]{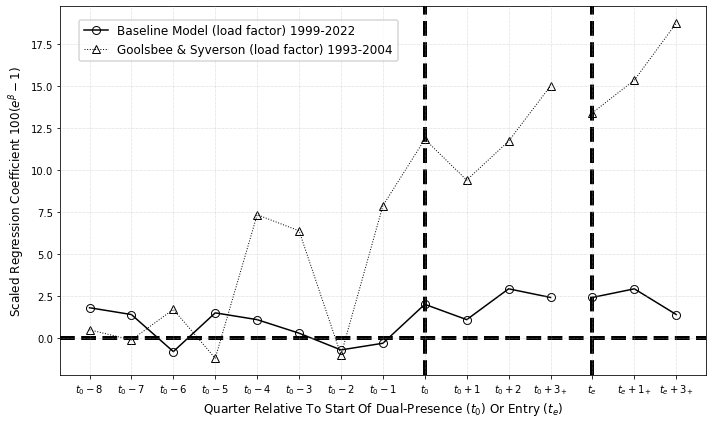}
        \caption{Percentage change in load factor (derived from T-100 passengers and available seats) relative to threat and entry events. This figure shows the standardized point estimates $100(e^{\beta}-1)$ of quarterly dummies relative to the start of Southwest's dual-presence at $t_{0}$ and actual entry at $t_{e}$. The solid line gives estimated percentage changes from my baseline model (\ref{eq:baseline}) that are computed from the point estimates $\beta$ reported in column 3 of Table \ref{tab:baseline_model_T100_robust}. For comparison, I include estimated percentage changes from column 4 of Goolsbee and Syverson's (2008, Table V) baseline model.}
        \label{fig:WN_regression_coefficients_model_1_loadfactor}
\end{figure}

\newpage
\noindent \emph{Evidence for a capacity reaction}\\
\\
I report the results for logged total passengers and logged available seats and the logged load factor (the ratio of passengers to seats), all using T-100 data, in  Table \ref{tab:baseline_model_T100_clustered} (clustered) and Table \ref{tab:baseline_model_T100_robust} (robust). I plot standardized point estimates from these models in Fig. \ref{fig:WN_regression_coefficients_model_1_passengers_seats} (passengers and seats) and Fig. \ref{fig:WN_regression_coefficients_model_1_loadfactor} (load factor). The results are striking! Contrary to Goolsbee and Syverson (2008)\nocite{goolsbee_syverson08}, I find evidence for substantial preemptive and post-entry capacity reactions that start up to six quarters before dual-presence, and that are matched by an increase in the number of passengers; both effects are highly significant in all specifications. The total number of passengers increases (pre-threat) from 10--37\% between $t_{0} - 6$ and $t_{0} - 1$, (post-threat) from 28--40\% between $t_{0}$ and $t_{0} + 12$, and (post-entry) from 27--46\% after $t_{e}$, all relative to the excluded period. While Goolsbee and Syverson's\nocite{goolsbee_syverson08} point estimates also seem to suggest, in particular, an increase in passenger numbers, their estimates of passenger and capacity changes are not significant. On the other hand, they report significant post-threat and post-entry increases of 9--16\% in load factor. However, I find no evidence that load factor changes significantly: at most, the point estimates indicate a 1--3\% increase after $t_{0}$. I argue that incumbents use capacity (through the number of available seats on a route) as a strong preemptive and post-entry strategic measure. Since capacity increases are less reversible than price cuts, this may be perceived by a potential entrant as a more credible entry deterrent.

\section{The Role of Network Structure}
I now examine whether an incumbent's price and capacity reactions are influenced by its global or local network structure. To do this, I modify my baseline model (\ref{eq:baseline}) with interactions as follows:
\begin{dmath}\label{eq:baseline_network}
\ln \, y_{\, ij \, t} = \left(\sum_{k \in \{-8, \ldots, -1, 0, 1, 2, 3+\}}\gamma_{\, k} \cdot D_{t_{0}(j) + k}\right) + \left(\sum_{k \in \{0, 1, 2, 3+\}}\psi_{\, k} \cdot D_{t_{0}(j) + k} \cdot Z_{ijt}\right) + \left(\sum_{k \in \{0, 1+, 3+\}}\gamma_{\, k} \cdot D_{t_{e}(j) + k}\right) + \left(\sum_{k \in \{0, 1+, 3+\}}\psi_{\, k} \cdot D_{t_{e}(j) + k} \cdot Z_{ijt}\right) + \alpha_{\, ij} + \delta_{\, t} + \beta \cdot X_{\, ij \, t} + \epsilon_{\, ij \, t},
\end{dmath}
where the post-threat and post-entry coefficients $\psi_{k}$ capture the interaction of time periods relative to dual-presence and entry with a single network variable $Z_{ijt}$. The approximate percentage change in $y$ from being in one of these time periods is now $100(\exp(\gamma_{k} + \psi_{k} \cdot Z_{ijt})-1)$. I include six global (carrier-quarter) network measures for $Z_{ijt}$: an incumbent's network density, diameter, average path length (a.p.l.), transitivity, average clustering coefficient, and (degree) assortativity. I also include six local (route-carrier-quarter) network measures for $Z_{ijt}$: a route's edge betweenness centrality; the maximum across the two endpoint nodes on a route of each of node degree centrality, node closeness centrality, node betweenness centrality, and node eigenvector centrality; and the maximum across both endpoint nodes on a route of the average degree of the node's neighbours.

\subsection{Network measures}
A \emph{graph} is an ordered pair $G = (V, E)$ where $V$ and $E$ denote the sets of \emph{nodes} and \emph{edges} of $G$, respectively. I use $n=|V|$ and $m=|E|$ to represent the numbers of nodes and edges. A graph is described by an $n \times n$ \emph{adjacency matrix} $g$, with representative element $(g)_{ij}$ which determines the nature of the link between nodes $i$ and $j$. A graph is \emph{simple and unweighted} if $(g)_{ii} = 0$ (no self-links) and $(g)_{ij} \in \{0, 1\}$ (no pair of nodes is linked by more than one edge, or by an edge with a weight that is different from one), and a graph is \emph{undirected} if $(g)_{ij} = (g)_{ji}$. I also use $(i, j) \in E$ to represent an edge between nodes $i$ and $j$, and say that they are \emph{directly-connected}. The neighbourhood of node $i$, denoted by $N(i) = \{ j : (i, j) \in E\}$, is the set of all nodes that are directly-connected to node $i$, and the \emph{degree} $k_{i} = \sum_{j}(g)_{ij}$ is the number of elements in $N(i)$. A \emph{walk} between nodes $i$ and $j$ is a sequence of edges $\{(i_{r}, i_{r+1})\}_{r=1,\ldots,R}$ such that $i_{1} = i$ and $i_{R+1} = j$, and a \emph{path} is a walk with distinct nodes. A graph is \emph{connected} if there is at least one path between any pair of nodes $i$ and $j$. In this paper, I only consider simple, unweighted, undirected and connected graphs. The matrix $g^{r} = (g^{r})_{ij}$ gives the number of walks of length $r$ between nodes $i$ and $j$. A \emph{geodesic} is any shortest path (least number of edges) between nodes $i$ and $j$ and has length $\ell_{ij}$. The total number of geodesics between nodes $a$ and $b$ is denoted $P(a, b)$. Then, $P_{i}(a, b)$ is the number of geodesics between nodes $a$ and $b$ that include node $i$, and $P_{(i, j)}(a, b)$ is the number of geodesics between nodes $a$ and $b$ that include edge $(i, j)$. I now define the six global and six local network measures that are included in (\ref{eq:baseline_network}), and demonstrate their calculation on a small illustrative network in Appendix \ref{sec:network_illustration}.\footnote{For each of the local node measures, I take the maximum over the two endpoints of the edge.}\\
\\
\noindent{\emph{Density}}\\
\\
The density is the number of edges $m$ relative to the maximum number of edges in a graph with $n$ nodes, $\binom{n}{2}$:
\begin{equation*}
    \mathrm{density} = \frac{2m}{n(n-1)}.
\end{equation*}
\noindent{\emph{Diameter}}\\
\\
The diameter is the maximum shortest path length across all pairs of nodes:
\begin{equation*}
    \mathrm{diameter} = \max_{i < j} \, \ell_{ij}.
\end{equation*}
\noindent{\emph{Average path length}}\\
\\
The average path length (a.p.l.) is the mean shortest path length across all pairs of nodes:
\begin{equation*}
    \mathrm{a.p.l.} = \frac{2}{n(n-1)}\sum_{i < j} \, \ell_{ij}.
\end{equation*}
\noindent{\emph{Transitivity}}\\
\\
The transitivity is the fraction of connected triples (three distinct nodes) that have their third edge connected to form a triangle:
\begin{equation*}
    \mathrm{transitivity} = \frac{\mathrm{tr}(g^{3})}{\sum_{i=1}^{n}k_{i}(k_{i}-1)}.
\end{equation*}
where $\mathrm{tr}(\cdot)$ is the trace of a square matrix.\\
\\
\noindent{\emph{Average clustering coefficient}}\\
\\
The average clustering coefficient (av. clustering) is the fraction of connected triples adjacent to node $i$ that are themselves connected, averaged over all nodes $i$:
\begin{equation*}
    \mathrm{av. \, clustering} = \frac{1}{n}\left\{\sum_{i=1}^{n}\frac{(g^{3})_{ii}}{k_{i}(k_{i}-1)}\right\}.
\end{equation*}
\noindent{\emph{Degree assortativity}}\\
\\
The degree assortativity coefficient (assortativity) is the Pearson correlation coefficient of the \emph{excess degree} $k - 1$ between directly-connected nodes (Newman, 2002\nocite{newman02}):
\begin{equation*}
    \mathrm{assortativity} = \frac{\frac{1}{m}\sum_{(i, j) \in E}(k_{i} - 1)(k_{j} - 1) - \left[\frac{1}{m}\sum_{(i, j) \in E}\frac{1}{2}((k_{i} - 1) + (k_{j} - 1))\right]^{2}}{\frac{1}{m}\sum_{(i, j) \in E}\frac{1}{2}((k_{i} - 1)^{2} + (k_{j} - 1)^{2}) - \left[\frac{1}{m}\sum_{(i, j) \in E}\frac{1}{2}((k_{i} - 1) + (k_{j} - 1))\right]^{2}}.
\end{equation*}
\noindent{\emph{Edge betweenness centrality}}\\
\\
The edge betweenness centrality of edge $(i, j)$ is:
\begin{equation*}
    \mathrm{edge \, between.} = \frac{2}{n(n-1)}\sum_{a < b}\frac{P_{(i, j)}(a, b)}{P(a, b)}.
\end{equation*}
\noindent{\emph{Degree centrality}}\\
\\
The degree centrality of node $i$ is:
\begin{equation*}
    \mathrm{degree} = \frac{k_{i}}{n-1}.
\end{equation*}
\noindent{\emph{Closeness centrality}}\\
\\
The closeness centrality of node $i$ is:
\begin{equation*}
    \mathrm{closeness} = \frac{n - 1}{\sum_{j \neq i}\ell_{ij}}.
\end{equation*}
\noindent{\emph{Betweenness centrality}}\\
\\
The betweenness centrality of node $i$ is:
\begin{equation*}
    \mathrm{between.} = \frac{2}{(n-1)(n-2)}\sum_{\substack{a < b \\ i \neq a, b}}\frac{P_{i}(a, b)}{P(a, b)}.
\end{equation*}
\noindent{\emph{Eigenvector centrality}}\\
\\
The eigenvector centrality of node $i$ is:
\begin{equation*}
    \mathrm{eigenv.} = (x)_{i},
\end{equation*}
where $x = (x)_{i}$ is the right-hand eigenvector that corresponds to the principal eigenvalue $\lambda$ in:
\begin{equation*}
    \lambda \, x = g \, x.
\end{equation*}
\noindent{\emph{Average neighbour degree}}\\
\\
The average neighbour degree of node $i$ is:
\begin{equation*}
    \mathrm{neighbour \, degree} = \frac{1}{k_{i}}\sum_{j \in N(i)}k_{j}.
\end{equation*}

\begin{table}
    \centering
    \begin{threeparttable}
    \caption{Incumbent Response to Dual-Presence and Entry by Southwest (clustered standard errors).}
     \label{tab:baseline_model_clustered_ses}
        \begin{tabular}{lcccccc}
            \hline\hline
            \rule{0pt}{2.5ex}
             & (1) & (2) & (3) & (4) & (5) & (6)\\
             & $\mathrm{ln}(P_{\mathrm{mean}})$ & $\mathrm{ln}(P_{10})$ & $\mathrm{ln}(P_{25})$ & $\mathrm{ln}(P_{75})$ & $\mathrm{ln}(P_{90})$ & $\mathrm{ln}(P_{\mathrm{mean}})$\\
            \midrule
            Before Dual-Presence & $0.016$ & $0.024$ & $0.032$ & $0.031$ & $-0.007$ & $0.012$ \\
            $t_{0} - 8$ & $(0.026)$ & $(0.028)$ & $(0.028)$ & $(0.029)$ & $(0.026)$ & $(0.026)$ \\
            Before Dual-Presence & $-0.004$ & $0.030$ & $0.010$ & $-0.013$ & $-0.009$ & $-0.007$ \\
            $t_{0} - 7$ & $(0.029)$ & $(0.037)$ & $(0.030)$ & $(0.027)$ & $(0.033)$ & $(0.028)$ \\
            Before Dual-Presence & $-0.030$ & $-0.018$ & $-0.046$ & $-0.042$ & $0.001$ & $-0.033$ \\
            $t_{0} - 6$ & $(0.032)$ & $(0.047)$ & $(0.037)$ & $(0.035)$ & $(0.031)$ & $(0.032)$ \\
            Before Dual-Presence & $-0.021$ & $-0.006$ & $-0.025$ & $-0.049$ & $-0.018$ & $-0.022$ \\
            $t_{0} - 5$ & $(0.038)$ & $(0.061)$ & $(0.050)$ & $(0.039)$ & $(0.034)$ & $(0.038)$ \\
            Before Dual-Presence & $0.009$ & $0.042$ & $0.020$ & $-0.007$ & $0.007$ & $0.010$ \\
            $t_{0} - 4$ & $(0.038)$ & $(0.049)$ & $(0.047)$ & $(0.041)$ & $(0.037)$ & $(0.037)$ \\
            Before Dual-Presence & $-0.004$ & $0.011$ & $-0.010$ & $-0.029$ & $0.010$ & $-0.0004$ \\
            $t_{0} - 3$ & $(0.038)$ & $(0.048)$ & $(0.047)$ & $(0.044)$ & $(0.037)$ & $(0.038)$ \\
            Before Dual-Presence & $0.010$ & $0.020$ & $-0.001$ & $0.0002$ & $0.045$ & $0.015$ \\
            $t_{0} - 2$ & $(0.040)$ & $(0.051)$ & $(0.051)$ & $(0.044)$ & $(0.037)$ & $(0.040)$ \\
            Before Dual-Presence & $0.006$& $0.049$ & $0.028$ & $-0.015$ & $-0.005$ & $0.010$ \\
            $t_{0} - 1$ & $(0.041)$ & $(0.063)$ & $(0.057)$ & $(0.047)$ & $(0.036)$ & $(0.042)$ \\ \hline
            Start of Dual-Presence & $-0.033$ & $-0.035$ & $-0.016$ & $-0.044$ & $-0.025$ & $-0.018$ \\
            $t_{0}$ & $(0.048)$ & $(0.068)$ & $(0.063)$ & $(0.051)$ & $(0.041)$ & $(0.050)$ \\
            During Dual-Presence & $-0.066$ & $-0.023$ & $-0.030$ & $-0.069$ & $-0.079^{*}$ & $-0.053$ \\
            $t_{0} + 1$ & $(0.052)$ & $(0.069)$ & $(0.068)$ & $(0.052)$ & $(0.042)$ & $(0.053)$ \\
            During Dual-Presence & $-0.083$ & $-0.060$ & $-0.063$ & $-0.111^{*}$ & $-0.064$ & $-0.075$ \\
            $t_{0} + 2$ & $(0.055)$ & $(0.079)$ & $(0.071)$ & $(0.058)$ & $(0.047)$ & $(0.055)$ \\
            During Dual-Presence & $-0.043$ & $0.005$ & $-0.012$ & $-0.057$ & $-0.063$ & $-0.036$ \\
            $t_{0} + 3$ to $t_{0} + 12$ & $(0.062)$ & $(0.089)$ & $(0.081)$ & $(0.063)$ & $(0.048)$ & $(0.063)$ \\ \hline
            Entry & $-0.177^{***}$ & $-0.206^{**}$ & $-0.200^{**}$ & $-0.156^{**}$ & $-0.172^{***}$ & $-0.165^{**}$ \\
            $t_{e}$ & $(0.063)$ & $(0.097)$ & $(0.088)$ & $(0.062)$ & $(0.047)$ & $(0.064)$ \\
            After Entry & $-0.194^{***}$ & $-0.190^{*}$ & $-0.203^{**}$ & $-0.183^{**}$ & $-0.191^{***}$ & $-0.181^{**}$ \\
            $t_{e} + 1$ to $t_{e} + 2$ & $(0.070)$ & $(0.103)$ & $(0.095)$ & $(0.074)$ & $(0.051)$ & $(0.071)$ \\
            After Entry & $-0.197^{**}$ & $-0.166$ & $-0.188^{*}$ & $-0.191^{**}$ & $-0.210^{***}$ & $-0.187^{**}$ \\
            $t_{e} + 3$ to $t_{e} + 12$ & $(0.082)$ & $(0.114)$ & $(0.108)$ & $(0.088)$ & $(0.062)$ & $(0.082)$ \\ \hline
            Distance (100 miles) & & & & & & $14.826^{**}$ \\
             & & & & & & $(6.103)$ \\
            Distance-squared & & & & & & $-0.832^{***}$ \\
             & & & & & & $(0.313)$ \\
            Temp. differential ($^{\circ}$F) & & & & & & $-0.003$ \\
             & & & & & & $(0.002)$ \\ \hline
            Carrier-route/time FE & Yes & Yes & Yes & Yes & Yes & Yes \\
            $R^{2}$ & $0.840$ & $0.814$ & $0.803$ & $0.800$ & $0.825$ & $0.843$ \\
            $N$ & $3,157$ & $3,157$ & $3,157$ & $3,157$ & $3,157$ & $3,157$ \\
            \hline\hline
        \end{tabular}
        \begin{tablenotes}[flushleft]
            \item \textit{Note.} The dependent variable is the passenger-weighted logged mean real fare or the 10th, 25th, 75th or 90th real fare percentile. I report standard errors clustered by carrier-route in parentheses. The asterisks $*$, $**$ and $***$ denote significance at the 90\%, 95\% and 99\% levels.
        \end{tablenotes}
    \end{threeparttable}
\end{table}

\clearpage

\begin{figure}[t]
        \centering
        \includegraphics[width=.75\linewidth, keepaspectratio]{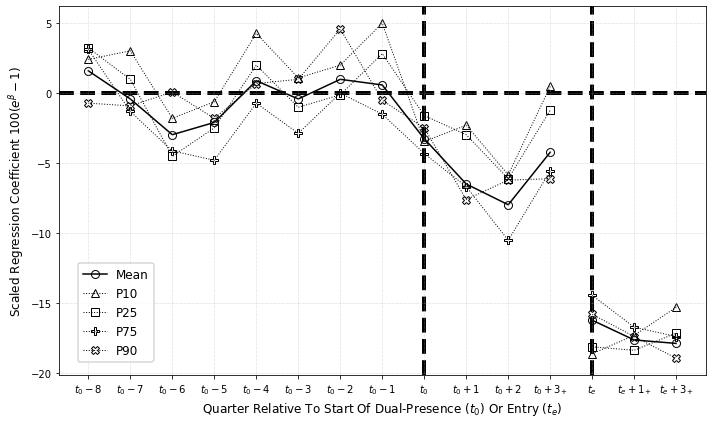}
        \caption{Percentage change in real fare percentiles relative to threat and entry events. This figure shows the standardized point estimates $100(e^{\beta}-1)$ of quarterly dummies relative to the start of Southwest's dual-presence at $t_{0}$ and actual entry at $t_{e}$. The estimated percentage changes from my baseline model (\ref{eq:baseline}) for each percentile are computed from the point estimates $\beta$ reported in columns 2--5 of Table \ref{tab:baseline_model}, with the mean fare effect included for comparison.}
        \label{fig:WN_regression_coefficients_model_1_percentiles}
\end{figure}

\clearpage

\begin{table}
    \centering
    \begin{threeparttable}
     \caption{Incumbent Response to Dual-Presence and Entry by Southwest: passengers, seat capacity and load factor (clustered standard errors).}
     \label{tab:baseline_model_T100_clustered}
        \begin{tabular}{lccc}
            \hline\hline
            \rule{0pt}{2.5ex}
             & (1) & (2) & (3)\\
             & $\mathrm{ln}(q)$ & $\mathrm{ln}(\mathrm{seats})$ & $\mathrm{ln}(\mathrm{load})$ \\
            \midrule
            Before Dual-Presence & $0.013$ & $-0.005$ & $0.018$\\
            $t_{0} - 8$ & $(0.051)$ & $(0.049)$ & $(0.011)$\\
            Before Dual-Presence & $0.055$ & $0.041$ & $0.014$\\
            $t_{0} - 7$ & $(0.045)$ & $(0.044)$ & $(0.011)$ \\
            Before Dual-Presence & $0.097^{**}$ & $0.105^{**}$ & $-0.008$ \\
            $t_{0} - 6$ & $(0.048)$ & $(0.047)$ & $(0.012)$ \\
            Before Dual-Presence & $0.195^{***}$ & $0.180^{***}$ & $0.015$  \\
            $t_{0} - 5$ & $(0.054)$ & $(0.050)$ & $(0.018)$ \\
            Before Dual-Presence & $0.195^{***}$ & $0.184^{***}$ & $0.011$\\
            $t_{0} - 4$ & $(0.057)$ & $(0.052)$ & $(0.016)$ \\
            Before Dual-Presence & $0.215^{***}$ & $0.211^{***}$ & $0.003$\\
            $t_{0} - 3$ & $(0.058)$ & $(0.056)$ & $(0.014)$\\
            Before Dual-Presence & $0.167^{**}$ & $0.174^{***}$ & $-0.007$\\
            $t_{0} - 2$ & $(0.067)$ & $(0.066)$ & $(0.022)$\\
            Before Dual-Presence & $0.312^{***}$ & $0.315^{***}$ & $-0.003$ \\
            $t_{0} - 1$ & $(0.106)$ & $(0.106)$ & $(0.018)$ \\ \hline
            Start of Dual-Presence & $0.282^{***}$& $0.262^{***}$ & $0.020$ \\
            $t_{0}$ & $(0.083)$ & $(0.080)$ & $(0.019)$\\
            During Dual-Presence & $0.251^{***}$ & $0.240^{***}$ & $0.011$\\
            $t_{0} + 1$ & $(0.080)$ & $(0.077)$ & $(0.016)$\\
            During Dual-Presence & $0.268^{**}$ & $0.239^{**}$ & $0.029^{*}$\\
            $t_{0} + 2$ & $(0.114)$ & $(0.112)$ & $(0.016)$\\
            During Dual-Presence & $0.339^{**}$ & $0.315^{***}$ & $0.024$\\
            $t_{0} + 3$ to $t_{0} + 12$ & $(0.104)$ & $(0.102)$ & $(0.017)$\\ \hline
            Entry & $0.380^{***}$ & $0.356^{***}$ & $0.024$\\
            $t_{e}$ & $(0.114)$ & $(0.106)$ & $(0.020)$\\
            After Entry & $0.368^{***}$ & $0.339^{***}$ & $0.029$\\
            $t_{e} + 1$ to $t_{e} + 2$ & $(0.117)$ & $(0.109)$ & $(0.020)$\\
            After Entry & $0.240^{*}$ & $0.226^{*}$ & $0.014$\\
            $t_{e} + 3$ to $t_{e} + 12$ & $(0.122)$ & $(0.114)$ & $(0.024)$\\ \hline
            Carrier-route/time FE & Yes & Yes & Yes\\
            $R^{2}$ & $0.891$ & $0.894$ & $0.656$\\
            $N$ & $3,157$ & $3,157$ & $3,157$ \\
            \hline\hline
        \end{tabular}
        \begin{tablenotes}[flushleft]
            \item \textit{Note.} The dependent variable is logged total passengers ($q$) or logged available seats (seats) from  T-100, or logged load factor (ratio of $q$ to seats). I report standard errors clustered by carrier-route in parentheses. The asterisks $*$, $**$ and $***$ denote significance at the 90\%, 95\% and 99\% levels.
        \end{tablenotes}
    \end{threeparttable}
\end{table}

\clearpage

\clearpage

\noindent \emph{Evidence for the role of network structure}\\
\\
I report results on logged mean fares from equation (\ref{eq:baseline_network}), with global and local network measures, in Tables \ref{tab:baseline_model_global_network_clustered_ses}--\ref{tab:baseline_model_local_network_clustered_ses} (clustered) and Tables \ref{tab:baseline_model_global_network}--\ref{tab:baseline_model_local_network} (robust). Post-entry fare cutting is influenced significantly by an incumbent's global network structure. Incumbents with a higher network density (or a lower diameter or average path length) cut fares by less, although there is no clear pattern of timing of the effects. For example, an incumbent with density 5\% (15\%) will cut fares by 20\% (12\%) between $t_{e} + 1$ and $t_{e} + 2$. There is weak evidence that global network effects matter preemptively too, if dual-presence continues for three quarters or more without entry. Incumbents with lower diameter (or average path length, average clustering or degree assortativity) cut fares by less. For example, an incumbent with average clustering coefficient 40\% (60\%) will cut fares by 2\% (4\%) between $t_{0} + 3$ and $t_{0} + 12$. There is some evidence that local network structure matters preemptively, if dual-presence lasts beyond three quarters. Incumbents cut fares by less on routes that are more central (edge betweenness and each of the maximum node centralities). For example, incumbents will cut fares by 10\% (3\%) on a route with (maximum) node closeness centrality of 60\% (90\%) between $t_{0} + 3$ and $t_{0} + 12$. There is no evidence for significant post-entry local structure interactions. One possible interpretation is that fare cuts are used less by firms with more ``efficient'' route networks (higher density, lower diameter or average path length), and on routes that occupy a more ``central'' position in an incumbent's network, in the sense that one of the endpoint airports is a hub for that carrier.

I report results on logged passengers and logged available seats in Tables \ref{tab:baseline_model_T100_global_network_clustered_ses}--\ref{tab:baseline_model_T100_seats_local_network_clustered_ses} (clustered) and Tables \ref{tab:baseline_model_T100_global_network}--\ref{tab:baseline_model_T100_seats_local_network} (robust). Using passengers or seats gives similar results, and so I focus on capacity. There is little statistical evidence that global or network structure matter preemptively. However, most of the network measures have post-entry interactions that are highly significant in all specifications, and the signs of the effects are constant over time. Capacity increases by less when an incumbent has a more efficient network structure (lower diameter or average path length), when there is more clustering (higher transitivity or average clustering coefficient), or when there is higher degree assortativity. For example, available seats increase by 29\% (13\%) for transitivity of 10\% (20\%) at $t_{e}$. On the other hand, capacity increases generally grow with a route's centrality. For example, available seats increase by 38\% (53\%) for edge betweenness of 2\% (4\%) at $t_{e}$.

\begin{table}
    \centering
    \begin{threeparttable}
     \caption{Incumbent Response to Dual-Presence and Entry by Southwest: global network measures (clustered ses).}     \label{tab:baseline_model_global_network_clustered_ses}
\begin{tabular}{lcccccc}
            \hline\hline
            \rule{0pt}{2.5ex}
             & (1) & (2) & (3) & (4) & (5) & (6)\\
             & $\mathrm{ln}(P_{\mathrm{mean}})$ & $\mathrm{ln}(P_{\mathrm{mean}})$ & $\mathrm{ln}(P_{\mathrm{mean}})$ & $\mathrm{ln}(P_{\mathrm{mean}})$ & $\mathrm{ln}(P_{\mathrm{mean}})$ & $\mathrm{ln}(P_{\mathrm{mean}})$\\
            network measure: & density & diameter & a.p.l. & transitivity & av. clustering & assortativity\\
            \midrule
            Start of Dual-Presence & $-0.041$ & $0.107^{*}$ & $0.577^{***}$ & $-0.036$ & $-0.017$ & $-0.026$ \\
            $t_{0}$ & $(0.071)$ & $(0.054)$ & $(0.107)$ & $(0.054)$ & $(0.049)$ & $(0.087)$ \\
            $\times$ network measure & $0.117$ & $-0.049^{***}$ & $-0.300^{***}$ & $0.035$ & $0.004$ & $0.013$\\
            & $(0.630)$ & $(0.014)$ & $(0.050)$ & $(0.217)$ & $(0.057)$ & $(0.094)$ \\
            During Dual-Presence & $-0.081$ & $0.043$ & $0.345^{**}$ & $-0.069$ & $-0.058$ & $-0.010$ \\
\hline\hline
        \end{tabular}
        \begin{tablenotes}[flushleft]
            \item \textit{Note.} The dependent variable is passenger-weighted logged mean real fare. I report standard errors clustered by carrier-route in parentheses. The asterisks $*$, $**$ and $***$ denote significance at the 90\%, 95\% and 99\% levels.
        \end{tablenotes}
    \end{threeparttable}
\end{table}

\begin{table}
    \centering
    \begin{threeparttable}
     \caption{Incumbent Response to Dual-Presence and Entry by Southwest: local network measures (clustered standard errors).}
     \label{tab:baseline_model_local_network_clustered_ses}
        \begin{tabular}{lcccccc}
            \hline\hline
            \rule{0pt}{2.5ex}
             & (1) & (2) & (3) & (4) & (5) & (6)\\
             & $\mathrm{ln}(P_{\mathrm{mean}})$ & $\mathrm{ln}(P_{\mathrm{mean}})$ & $\mathrm{ln}(P_{\mathrm{mean}})$ & $\mathrm{ln}(P_{\mathrm{mean}})$ & $\mathrm{ln}(P_{\mathrm{mean}})$ & $\mathrm{ln}(P_{\mathrm{mean}})$\\
            network measure: & edge between. & degree & closeness & between. & eigenv. & neighbour degree\\
            \midrule
            Start of Dual-Presence & $-0.047$ & $-0.153^{**}$ & $-0.267^{***}$ & $-0.066$ & $-0.075$ & $-0.067$ \\
            $t_{0}$ & $(0.050)$ & $(0.067)$ & $(0.098)$ & $(0.054)$ & $(0.062)$ & $(0.063)$ \\
            $\times$ network measure & $0.476$ & $0.153^{**}$ & $0.270^{***}$ & $0.067$ & $0.119$ & $0.001$\\
            & $(0.413)$ & $(0.057)$ & $(0.094)$ & $(0.042)$ & $(0.092)$ & $(0.001)$ \\
            During Dual-Presence & $-0.065$ & $-0.086$ & $-0.120$ & $-0.063$ & $-0.046$ & $-0.042$ \\
\hline\hline
        \end{tabular}
        \begin{tablenotes}[flushleft]
            \item \textit{Note.} The dependent variable is the passenger-weighted logged mean real fare. I report standard errors clustered by carrier-route in parentheses. The asterisks $*$, $**$ and $***$ denote significance at the 90\%, 95\% and 99\% levels.
        \end{tablenotes}
    \end{threeparttable}
\end{table}

\begin{table}
    \centering
    \begin{threeparttable}
     \caption{Incumbent Response to Dual-Presence and Entry by Southwest: global network measures (clustered std. errors).}
     \label{tab:baseline_model_T100_global_network_clustered_ses}
        \begin{tabular}{lcccccc}
            \hline\hline
            \rule{0pt}{2.5ex}
             & (1) & (2) & (3) & (4) & (5) & (6)\\
             & $\mathrm{ln}(q)$ & $\mathrm{ln}(q)$ & $\mathrm{ln}(q)$ & $\mathrm{ln}(q)$ & $\mathrm{ln}(q)$ & $\mathrm{ln}(q)$\\
            network measure: & density & diameter & a.p.l. & transitivity & av. clustering & assortativity\\
            \midrule
            Start of Dual-Presence & $0.174$ & $0.521^{***}$ & $0.841^{**}$ & $0.235^{***}$ & $0.294^{***}$ & $0.125$ \\     
            $t_{0}$ & $(0.112)$ & $(0.117)$ & $(0.393)$ & $(0.085)$ & $(0.083)$ & $(0.138)$ \\
            $\times$ network measure & $0.932$ & $-0.086^{***}$ & $-0.270$ & $-0.611$ & $-0.205^{*}$ & $-0.236$\\            
            & $(1.095)$ & $(0.032)$ & $(0.189)$ & $(0.385)$ & $(0.106)$ & $(0.179)$ \\
            During Dual-Presence & $0.409^{***}$ & $0.384^{*}$ & $-0.340$ & $0.224^{**}$ & $0.283^{***}$ & $0.233$ \\
\hline\hline
        \end{tabular}
        \begin{tablenotes}[flushleft]
            \item \textit{Note.} The dependent variable is the number of passengers (T-100). I report standard errors clustered by carrier-route in parentheses. The asterisks $*$, $**$ and $***$ denote significance at the 90\%, 95\% and 99\% levels.
        \end{tablenotes}
    \end{threeparttable}
\end{table}

\begin{table}
    \centering
    \begin{threeparttable}
     \caption{Incumbent Response to Dual-Presence and Entry by Southwest: local network measures (clustered std. errors).}
     \label{tab:baseline_model_T100_local_network_clustered_ses}
        \begin{tabular}{lcccccc}
            \hline\hline
            \rule{0pt}{2.5ex}
             & (1) & (2) & (3) & (4) & (5) & (6)\\
             & $\mathrm{ln}(q)$ & $\mathrm{ln}(q)$ & $\mathrm{ln}(q)$ & $\mathrm{ln}(q)$ & $\mathrm{ln}(q)$ & $\mathrm{ln}(q)$\\
            network measure: & edge between. & degree & closeness & between. & eigenv. & neighbour degree\\
            \midrule
            Start of Dual-Presence & $0.198^{**}$ & $0.206^{*}$ & $0.103$ & $0.175^{*}$ & $0.137$ & $0.337^{***}$ \\     
            $t_{0}$ & $(0.082)$ & $(0.112)$ & $(0.153)$ & $(0.100)$ & $(0.115)$ & $(0.084)$ \\                 
            $\times$ network measure & $1.941^{**}$ & $0.010$ & $0.103$ & $0.069$ & $0.252$ & $-0.002$\\            
            & $(0.779)$ & $(0.114)$ & $(0.169)$ & $(0.085)$ & $(0.187)$ & $(0.002)$ \\
            During Dual-Presence & $0.251^{***}$ & $0.225$ & $0.122$ & $0.143$ & $0.183$ & $0.144$ \\
\hline\hline
        \end{tabular}
        \begin{tablenotes}[flushleft]
            \item \textit{Note.} The dependent variable is the number of passengers (T-100). I report standard errors clustered by carrier-route in parentheses. The asterisks $*$, $**$ and $***$ denote significance at the 90\%, 95\% and 99\% levels.
        \end{tablenotes}
    \end{threeparttable}
\end{table}

\begin{table}
    \centering
    \begin{threeparttable}
     \caption{Incumbent Response to Dual-Presence and Entry by Southwest: global network measures (clustered std. errors).}
     \label{tab:baseline_model_T100_seats_global_network_clustered_ses}
        \begin{tabular}{lcccccc}
            \hline\hline
            \rule{0pt}{2.5ex}
             & (1) & (2) & (3) & (4) & (5) & (6)\\
             & $\mathrm{ln}(\mathrm{seats})$ & $\mathrm{ln}(\mathrm{seats})$ & $\mathrm{ln}(\mathrm{seats})$ & $\mathrm{ln}(\mathrm{seats})$ & $\mathrm{ln}(\mathrm{seats})$ & $\mathrm{ln}(\mathrm{seats})$\\
            network measure: & density & diameter & a.p.l. & transitivity & av. clustering & assortativity\\
            \midrule
            Start of Dual-Presence & $0.208$ & $0.434^{***}$ & $0.518$ & $0.213^{***}$ & $0.261^{***}$ & $0.172$ \\     
            $t_{0}$ & $(0.105)$ & $(0.115)$ & $(0.341)$ & $(0.080)$ & $(0.079)$ & $(0.132)$ \\
            $\times$ network measure & $0.328$ & $-0.063^{*}$ & $-0.123$ & $-0.487$ & $-0.164$ & $-0.138$\\            
            & $(1.019)$ & $(0.032)$ & $(0.163)$ & $(0.371)$ & $(0.104)$ & $(0.170)$ \\
            During Dual-Presence & $0.344^{***}$ & $0.280$ & $-0.362$ & $0.184^{**}$ & $0.233^{**}$ & $0.295^{**}$ \\
\hline\hline
        \end{tabular}
        \begin{tablenotes}[flushleft]
            \item \textit{Note.} The dependent variable is the total available seats (T-100). I report standard errors clustered by carrier-route in parentheses. The asterisks $*$, $**$ and $***$ denote significance at the 90\%, 95\% and 99\% levels.
        \end{tablenotes}
    \end{threeparttable}
\end{table}

\begin{table}
    \centering
    \begin{threeparttable}
     \caption{Incumbent Response to Dual-Presence and Entry by Southwest: local network measures (clustered std. errors).}
     \label{tab:baseline_model_T100_seats_local_network_clustered_ses}
        \begin{tabular}{lcccccc}
            \hline\hline
            \rule{0pt}{2.5ex}
             & (1) & (2) & (3) & (4) & (5) & (6)\\
             & $\mathrm{ln}(\mathrm{seats})$ & $\mathrm{ln}(\mathrm{seats})$ & $\mathrm{ln}(\mathrm{seats})$ & $\mathrm{ln}(\mathrm{seats})$ & $\mathrm{ln}(\mathrm{seats})$ & $\mathrm{ln}(\mathrm{seats})$\\
            network measure: & edge between. & degree & closeness & between. & eigenv. & neighbour degree\\
            \midrule
            Start of Dual-Presence & $0.202^{**}$ & $0.199^{*}$ & $0.125$ & $0.173^{*}$ & $0.150$ & $0.292^{***}$ \\     
            $t_{0}$ & $(0.078)$ & $(0.109)$ & $(0.153)$ & $(0.096)$ & $(0.113)$ & $(0.081)$ \\
            $\times$ network measure & $1.210$ & $0.002$ & $0.062$ & $0.055$ & $0.193$ & $-0.001$\\            
            & $(0.754)$ & $(0.109)$ & $(0.167)$ & $(0.083)$ & $(0.179)$ & $(0.002)$ \\
            During Dual-Presence & $0.246^{***}$ & $0.241^{*}$ & $0.183$ & $0.169^{*}$ & $0.225^{*}$ & $0.158^{*}$ \\
\hline\hline
        \end{tabular}
        \begin{tablenotes}[flushleft]
            \item \textit{Note.} The dependent variable is the total available seats (T-100). I report standard errors clustered by carrier-route in parentheses. The asterisks $*$, $**$ and $***$ denote significance at the 90\%, 95\% and 99\% levels.
        \end{tablenotes}
    \end{threeparttable}
\end{table}

\section{Conclusion}

I investigate how incumbent airlines in the U.S. respond to threatened and actual route entry by Southwest Airlines. Following Goolsbee and Syverson (2008)\nocite{goolsbee_syverson08}, I use a two-way fixed effects event study approach, but with the latest available data from 1999--2022, to identify a firm's price and quantity response. I find that incumbents cut fares preemptively (post-entry) by 6--8\% (16--18\%) although the significance, pattern, and timing of the preemptive cuts are quite different to Goolsbee and Syverson's\nocite{goolsbee_syverson08} earlier results. Incumbents increase capacity preemptively by 10--40\%, up to six quarters before the entry threat is established, and by 27--46\% post-entry. My results suggest a clear shift in firms' strategic response from price to quantity. I also investigate the impact of an incumbent's network structure on its preemptive and post-entry behaviour. Although the results on price are unclear, a firm's post-entry capacity reaction appears to depend on its global network structure as well as the local importance (centrality) of the route.

\clearpage

\singlespace

\bibliographystyle{chicago}

\bibliography{ms}

\newpage

\appendix

\counterwithin{figure}{section}

\section{Route Network and Supplementary Data}\label{sec:dataprocessing}
I construct the main dataset using two publicly-available sources over the years 1999 to 2022, collected from reporting carriers by the U.S. Department of Transportation’s Bureau of Transportation Statistics.  The ``Airline Origin and Destination Survey'', or DB1B, is a quarterly 10\% random sample of U.S. domestic airline itineraries. The DB1B is split into three datasets (Coupon, Ticket, and Market), which include operating and ticketing carriers, origin and destination airports, type of ticket, connections, number of passengers, and nominal dollar fares. I use the DB1B Coupon and DB1B Ticket data (the latter contains the itinerary-level fares). The raw datasets are extensive, e.g., the DB1B Coupon dataset has roughly 9 million individual itinerary segments (lines) in 2022Q1. There is no information in the aggregated data on the actual date of flight or purchase, or on ticket restrictions, or on the buyer’s characteristics. The ``T-100 Domestic Segment (All Carriers)'', or T-100, is a monthly 100\% census that contains traffic data on domestic nonstop flight segments, including the number of enplaned passengers and available capacity. The DB1B (and its predecessor the DB1A) and the T-100 have both been widely used in empirical work in economics (for example, studies of barriers to entry, entry threats, hub premia, and price dispersion, and structural models of market structure and profitability, by Goolsbee and Syverson, 2008\nocite{goolsbee_syverson08}; Ciliberto and Tamer, 2009\nocite{ciliberto_tamer09}; Gerardi and Shapiro, 2009\nocite{gerardi_shapiro09}; Berry and Jia, 2010\nocite{berry_jia10}; Ciliberto and Williams, 2010\nocite{ciliberto_williams10}; Aguirregabiria and Ho, 2012\nocite{aguirregabiria_ho12}; Cornia et al., 2012\nocite{cornia_etal12}; Dai et al., 2014\nocite{dai_etal14}; Snider and Williams, 2015\nocite{snider_williams15}). Both datasets are available at www.transtats.bts.gov. I filter, aggregate and merge the data from the DB1B Coupon, DB1B Ticket, and T-100, to the route-carrier-quarter level.

\subsection{Parsing the DB1B Coupon itinerary data}\label{sec:coupon_parse}
I retain a DB1B Coupon itinerary if all of the following conditions are satisified:
\begin{enumerate}
    \item (Nonstop Round-Trip) There are two segments (outbound and return) and two trip breaks on the itinerary, and the outbound origin and return destination airports are the same.
    
    \item (No Codesharing) The operating and ticketing carriers are the same on all segments (Gerardi and Shapiro, 2009\nocite{gerardi_shapiro09}; Dai et al., 2014\nocite{dai_etal14}).
    
    \item (Fare Class, Passengers) The itinerary is reported as (restricted or unrestricted) coach class, business class, or first class, and the fare class and the number of passengers are constant across the segments. The DB1B combines some identical tickets into a single itinerary code, and the appearance of multiple passengers on a given itinerary does not imply that the tickets were booked together, nor that the passengers travelled together.
    
    \item (Domestic Routes) The origin and destination airports are located in the continental U.S. (excluding Alaska).
    
    \item (No Cabotage) The operating carrier is a U.S. domestic airline. I exclude cabotage, which is air traffic that starts and ends within the boundaries of a given country by an air carrier of another country.
\end{enumerate}
In 2022Q1, there are 703,189 individual itineraries that satisfy these filters. For example, the itinerary code 202213184820 corresponds to a nonstop restricted coach class round-trip ticket, for one passenger, flying with United Airlines between Denver 
 (DEN) and Chicago O'Hare (ORD), a one-way distance of 888 miles.

\subsection{Parsing the DB1B Ticket itinerary data and merging it with the DB1B Coupon itinerary data}
I retain a DB1B Ticket itinerary if all of the following conditions are satisifed:
\begin{enumerate}
    \item (Credible Fare) I drop itineraries that have fares above the Bureau of Transportation Statistic's credible limit of \$1 per mile of travel (Gerardi and Shapiro, 2009\nocite{gerardi_shapiro09}; Cornia et al., 2012\nocite{cornia_etal12}; Dai et al., 2014\nocite{dai_etal14}).
    
    \item (Reliable Bulk Fare) Some multi-passenger ``bulk fare'' tickets are sold in large blocks of seats, and the carriers must divide the total number of passengers into the total dollar fee collected. I drop bulk fare itineraries for which the Bureau of Transportation Statistics considers the reported fare to be unreliable. This can capture complicated cases involving several departments within the airline, where the exact value is not known when the ticket is processed, and that could give an inaccurate fare.
\end{enumerate}
I then merge the filtered DB1B Coupon and DB1B Ticket data, retaining an itinerary only if it appears in both datasets, and if the value of any itinerary-level variables that are recorded in both Coupon and Ticket are the same. In 2022Q1, this leaves us with 702,274 individual itineraries. We can now see that itinerary 202213184820 from Denver to Chicago O'Hare was purchased for a nominal fare of \$518.

\subsection{Aggregating merged DB1B Coupon and Ticket itineraries to the route-carrier-quarter level}
I aggregate the individual itineraries to a unidirectional route-carrier-quarter level, preserving the fare distribution. In 2022Q1 there are 2,322 observations. I then apply the following filters:
\begin{enumerate}
    \item (Coach Class and Other Classes) If more than 75\% of a carrier’s tickets are recorded as business or first class, across all routes in a given quarter, then I retain all tickets for that carrier, regardless of the fare class; otherwise, I retain only coach class tickets. In each year from 1998 to 2013, some of Southwest Airlines' tickets are listed as first class; from 1998 to 2009, this number is considerable, e.g., in 2008, Southwest reported primarily first class tickets on 1,915 of its route-quarters. However, the fares for these tickets are typically quite low, e.g., in 2009Q1, Southwest reported 4,093 first class tickets, and no coach class tickets, on the route between Denver (DEN) and Phoenix Sky Harbor (PHX); the nominal fares range from \$70 to \$487, with a median fare of \$180, and a 90th percentile of \$283. A similar phenomenon has been reported for JetBlue.\footnote{The issue of Southwest’s fare class has also been noted by Ciliberto and Williams (2010, p.492)\nocite{ciliberto_williams10} for 1993--2005: ``Southwest codes all tickets under one fare class, despite selling tickets with different fare restrictions''; and by Dai et al., (2014, p.170, footnote 34)\nocite{dai_etal14} for 1993--2008: ``Since all itineraries issued by Southwest and JetBlue are reported as either first-class or business class, we code them as coach class.''} For this reason, I make no distinction between coach and business or first class tickets when a carrier reports an unusually high percentage of the latter.

    \item (Frequent-Flyer) The DB1B does not identify tickets that were purchased using frequent-flyer miles, or that were heavily discounted for airline employees. I try to exclude both, by eliminating all round-trip tickets that have a nominal fare below \$20 (Gerardi and Shapiro, 2009\nocite{gerardi_shapiro09}; Cornia et al., 2012\nocite{cornia_etal12}; Dai et al., 2014\nocite{dai_etal14}; Tan, 2016\nocite{tan16}; while Brueckner and Whalen, 2000,\nocite{brueckner_whalen00} remove round-trip fares that are below \$100).

    \item (DB1B Low Volume) I remove all route-carrier-quarters with fewer than 100 passengers in the DB1B (a similar condition is imposed by Armantier and Richard, 2008\nocite{armantier_richard08}; Gerardi and Shapiro, 2009\nocite{gerardi_shapiro09}; while Ciliberto and Williams, 2010,\nocite{ciliberto_williams10} drop route-carriers with less than 20 passengers, to exclude regional service and smaller planes). This corresponds to about 2 to 6 full planes per quarter, or roughly 1 full plane per fortnight.

     \item (Very High Fares) To catch possible data recording errors, I remove tickets for which the nominal fare exceeds the 99th percentile of the route-carrier-quarter fare distribution (Dai et al., 2014\nocite{dai_etal14}).

     \item (Southwest at Dallas/Fort Worth International) Following an observation by Goolsbee and Syverson (2008)\nocite{goolsbee_syverson08}, I remove all Southwest route-quarters that involve Dallas/Fort Worth (DFW), from 1993Q1 to 1999Q4.\footnote{``We discovered [\ldots] an important airport error in the DB1A data. In several quarters, the DB1A source data indicate that Southwest operated flights out of Dallas/Ft. Worth Airport (DFW) in the late 1990s for a few quarters and then exited. Although the data show them flying from DFW to many different airports, the airline code for DFW must be mistaken. There is no record of Southwest operating these numerous flights out of DFW in the local business press at the time or in any other Department of Transportation data such as the T-100 [\ldots] We therefore dropped these DFW observations from the sample.'' (Goolsbee and Syverson, 2008, p.1615, footnote 6\nocite{goolsbee_syverson08})} (This condition is not binding: given the previous filters, there are no DFW routes for Southwest in the data.)

     \item (Small Network) So that I can compute meaningful carrier-quarter route network measures, I only retain carriers that serve at least ten routes in a given quarter.
\end{enumerate}
In 2022Q1, there are 2,623 route-carrier observations in the data. For example, a nonstop unidirectional round-trip from Washington Dulles (IAD) to Los Angeles (LAX), operated by United Airlines in 2022Q1, with 2,591 passengers recorded in the DB1B, is a single observation. The nominal fares for that route-carrier range from \$20 to \$1,697, with a median fare of \$381 and a 90th percentile of \$830. For all quarters, I compute real fares relative to 2022Q1 using monthly price index data from the U.S. Bureau of Labor Statistics (``CPI for All Urban Consumers (CPI-U)''). I created a quarterly deflator by taking the last month of each quarter. For example, 2012Q1\$300 = 2022Q1\$376, which represents a 25\% increase in prices between 2012Q1 and 2022Q1.

\subsection{Parsing T-100 route-carrier-month data, aggregating to quarters, and merging with the DB1B}\label{sec:T100parse}
A record (line) in the T-100 represents an aggregation of all flights to a directional route-carrier-month level, conditional on characteristics such as service class and aircraft type. In 2022Q1, there are 100,963 such records. I first apply a scheduled service filter:

\begin{enumerate}
    \item (Domestic Scheduled Flight) I retain a route-carrier-month only if it corresponds to domestic scheduled passenger/cargo service and the plane is configured for passengers or combined passengers and freight.
\end{enumerate}
In 2022Q1, this leaves 69,776 records. I then aggregate the T-100 to a unidirectional route-carrier-quarter level, collapsing all conditioning variables, and apply the following volume filter:

\begin{enumerate}
    \setcounter{enumi}{1}
    \item (T-100 Low Volume) I remove route-carrier-quarters with fewer than 2,000 enplaned passengers or 2,000 available seats. This is similar to the DB1B Low Volume filter, scaled up by a factor of 20: while the DB1B is a 10\% random sample, the T-100 contains all passengers; furthermore, the T-100 double-counts round-trip passengers, who appear once on the outbound segment, and again on the return segment.
\end{enumerate}
In 2022Q1, this data treatment gives 5,214 route-carrier observations in the T-100. I then merge the filtered and aggregated DB1B Coupon/Ticket and T-100 data, retaining a route-carrier-quarter observation only if it appears in both datasets. In 2022Q1, I keep 2,604 route-carrier observations, which is more than 99\% of the DB1B pre-merge route-carriers. In 2013Q4, there are more than two million passengers (in the 10\% sample), who travel with one of the following thirteen airlines (there are additional carriers in other quarters): AirTran Airways (FL); Alaska Airlines (AS); Allegiant Air (G4); American Airlines (AA); Delta Air Lines (DL); Frontier Airlines (F9); JetBlue Airways (B6); Southwest Airlines (WN); Spirit Airlines (NK); Sun Country Airlines (SY); United Airlines (UA); US Airways (US); Virgin America (VX). There are 1,931 distinct unidirectional airport-airport routes between 206 airports. The route distances range from 83 to 2,724 (statute) miles, with a mean distance of 1,021 miles.\\
\\
\noindent I repeat steps \ref{sec:coupon_parse}--\ref{sec:T100parse} for each of the 96 quarters from 1999--2022 to give the core route-carrier network data.

\subsection{Temperature data}
I supplement the DB1B and T-100 data with monthly state-level temperature, collected from the National Centers for Environmental Information (NCEI) of the U.S. Department of Commerce’s National Oceanic and Atmospheric Administration (NOAA).\footnote{I obtained state-level files from www.ncei.noaa.gov/access/monitoring/climate-at-a-glance/statewide/time-series, in September 2023.} I associate each airport in the sample with a single geographical state as reported in the DB1B. I then construct an annual route-level temperature variable, as the absolute difference in January average temperature (in Fahrenheit) between the origin and destination airport states.

\section{Robustness Checks}\label{sec:appendix_robustness}
In this appendix, I report regression results using robust standard errors.

\begin{table}
    \centering
    \begin{threeparttable}
     \caption{Incumbent Response to Dual-Presence and Entry by Southwest.}
     \label{tab:baseline_model}
        \begin{tabular}{lcccccc}
            \hline\hline
            \rule{0pt}{2.5ex}
             & (1) & (2) & (3) & (4) & (5) & (6)\\
             & $\mathrm{ln}(P_{\mathrm{mean}})$ & $\mathrm{ln}(P_{10})$ & $\mathrm{ln}(P_{25})$ & $\mathrm{ln}(P_{75})$ & $\mathrm{ln}(P_{90})$ & $\mathrm{ln}(P_{\mathrm{mean}})$\\
            \midrule
            Before Dual-Presence & $0.016$ & $0.024$ & $0.032$ & $0.031$ & $-0.007$ & $0.012$ \\
            $t_{0} - 8$ & $(0.019)$ & $(0.024)$ & $(0.022)$ & $(0.022)$ & $(0.021)$ & $(0.019)$ \\
            Before Dual-Presence & $-0.004$ & $0.030$ & $0.010$ & $-0.013$ & $-0.009$ & $-0.007$ \\
            $t_{0} - 7$ & $(0.020)$ & $(0.025)$ & $(0.024)$ & $(0.023)$ & $(0.022)$ & $(0.020)$ \\
            Before Dual-Presence & $-0.030$ & $-0.018$ & $-0.046^{*}$ & $-0.042^{*}$ & $0.001$ & $-0.033^{*}$ \\
            $t_{0} - 6$ & $(0.020)$ & $(0.025)$ & $(0.024)$ & $(0.023)$ & $(0.022)$ & $(0.020)$ \\
            Before Dual-Presence & $-0.021$ & $-0.006$ & $-0.025$ & $-0.049^{**}$ & $-0.018$ & $-0.022$ \\
            $t_{0} - 5$ & $(0.020)$ & $(0.025)$ & $(0.024)$ & $(0.023)$ & $(0.022)$ & $(0.020)$ \\
            Before Dual-Presence & $0.009$ & $0.042$ & $0.020$ & $-0.007$ & $0.007$ & $0.010$ \\
            $t_{0} - 4$ & $(0.020)$ & $(0.025)$ & $(0.024)$ & $(0.023)$ & $(0.022)$ & $(0.020)$ \\
            Before Dual-Presence & $-0.004$ & $0.011$ & $-0.010$ & $-0.029$ & $0.010$ & $-0.0004$ \\
            $t_{0} - 3$ & $(0.021)$ & $(0.026)$ & $(0.024)$ & $(0.024)$ & $(0.023)$ & $(0.020)$ \\
            Before Dual-Presence & $0.010$ & $0.020$ & $-0.001$ & $0.0002$ & $0.045^{**}$ & $0.015$ \\
            $t_{0} - 2$ & $(0.020)$ & $(0.025)$ & $(0.024)$ & $(0.023)$ & $(0.022)$ & $(0.020)$ \\
            Before Dual-Presence & $0.006$& $0.049^{*}$ & $0.028$ & $-0.015$ & $-0.005$ & $0.010$ \\
            $t_{0} - 1$ & $(0.021)$ & $(0.027)$ & $(0.026)$ & $(0.025)$ & $(0.024)$ & $(0.021)$ \\ \hline
            Start of Dual-Presence & $-0.033$ & $-0.035$ & $-0.016$ & $-0.044^{*}$ & $-0.025$ & $-0.018$ \\
            $t_{0}$ & $(0.021)$ & $(0.027)$ & $(0.025)$ & $(0.025)$ & $(0.023)$ & $(0.022)$ \\
            During Dual-Presence & $-0.066^{***}$ & $-0.023$ & $-0.030$ & $-0.069^{***}$ & $-0.079^{***}$ & $-0.053^{**}$ \\
            $t_{0} + 1$ & $(0.022)$ & $(0.027)$ & $(0.026)$ & $(0.026)$ & $(0.025)$ & $(0.022)$ \\
            During Dual-Presence & $-0.083^{***}$ & $-0.060^{**}$ & $-0.063^{**}$ & $-0.111^{***}$ & $-0.064^{**}$ & $-0.075^{***}$ \\
            $t_{0} + 2$ & $(0.022)$ & $(0.028)$ & $(0.027)$ & $(0.026)$ & $(0.025)$ & $(0.022)$ \\
            During Dual-Presence & $-0.043^{***}$ & $0.005$ & $-0.012$ & $-0.057^{***}$ & $-0.063^{***}$ & $-0.036^{**}$ \\
            $t_{0} + 3$ to $t_{0} + 12$ & $(0.017)$ & $(0.021)$ & $(0.020)$ & $(0.019)$ & $(0.018)$ & $(0.017)$ \\ \hline
            Entry & $-0.177^{***}$ & $-0.206^{***}$ & $-0.200^{***}$ & $-0.156^{***}$ & $-0.172^{***}$ & $-0.165^{***}$ \\
            $t_{e}$ & $(0.023)$ & $(0.029)$ & $(0.027)$ & $(0.027)$ & $(0.026)$ & $(0.023)$ \\
            After Entry & $-0.194^{***}$ & $-0.190^{***}$ & $-0.203^{***}$ & $-0.183^{***}$ & $-0.191^{***}$ & $-0.181^{***}$ \\
            $t_{e} + 1$ to $t_{e} + 2$ & $(0.021)$ & $(0.027)$ & $(0.025)$ & $(0.024)$ & $(0.023)$ & $(0.021)$ \\
            After Entry & $-0.197^{***}$ & $-0.166^{***}$ & $-0.188^{***}$ & $-0.191^{***}$ & $-0.210^{***}$ & $-0.187^{***}$ \\
            $t_{e} + 3$ to $t_{e} + 12$ & $(0.022)$ & $(0.028)$ & $(0.026)$ & $(0.025)$ & $(0.024)$ & $(0.022)$ \\ \hline
            Distance (100 miles) & & & & & & $14.826^{***}$ \\
             & & & & & & $(3.538)$ \\
            Distance-squared & & & & & & $-0.832^{***}$ \\
             & & & & & & $(0.148)$ \\
            Temp. differential ($^{\circ}$F) & & & & & & $-0.003^{***}$ \\
             & & & & & & $(0.001)$ \\ \hline
            Carrier-route/time FE & Yes & Yes & Yes & Yes & Yes & Yes \\
            $R^{2}$ & $0.840$ & $0.814$ & $0.803$ & $0.800$ & $0.825$ & $0.843$ \\
            $N$ & $3,157$ & $3,157$ & $3,157$ & $3,157$ & $3,157$ & $3,157$ \\
            \hline\hline
        \end{tabular}
        \begin{tablenotes}[flushleft]
            \item \textit{Note.} The dependent variable is the passenger-weighted logged mean real fare or the 10th, 25th, 75th or 90th real fare percentile. I report robust standard errors in parentheses. The asterisks $*$, $**$ and $***$ denote significance at the 90\%, 95\% and 99\% levels.
        \end{tablenotes}
    \end{threeparttable}
\end{table}

\begin{table}
    \centering
    \begin{threeparttable}
     \caption{Incumbent Response to Dual-Presence and Entry by Southwest: passengers, capacity and load factor.}
     \label{tab:baseline_model_T100_robust}
        \begin{tabular}{lccc}
            \hline\hline
            \rule{0pt}{2.5ex}
             & (1) & (2) & (3)\\
             & $\mathrm{ln}(q)$ & $\mathrm{ln}(\mathrm{seats})$ & $\mathrm{ln}(\mathrm{load})$ \\
            \midrule
            Before Dual-Presence & $0.013$ & $-0.005$ & $0.018^{*}$\\
            $t_{0} - 8$ & $(0.039)$ & $(0.037)$ & $(0.010)$\\
            Before Dual-Presence & $0.055$ & $0.041$ & $0.014$\\
            $t_{0} - 7$ & $(0.042)$ & $(0.040)$ & $(0.011)$ \\
            Before Dual-Presence & $0.097^{**}$ & $0.105^{***}$ & $-0.008$ \\
            $t_{0} - 6$ & $(0.041)$ & $(0.039)$ & $(0.011)$ \\
            Before Dual-Presence & $0.195^{***}$ & $0.180^{***}$ & $0.015$  \\
            $t_{0} - 5$ & $(0.042)$ & $(0.040)$ & $(0.011)$ \\
            Before Dual-Presence & $0.195^{***}$ & $0.184^{***}$ & $0.011$\\
            $t_{0} - 4$ & $(0.042)$ & $(0.040)$ & $(0.011)$ \\
            Before Dual-Presence & $0.215^{***}$ & $0.211^{***}$ & $0.003$\\
            $t_{0} - 3$ & $(0.043)$ & $(0.041)$ & $(0.011)$\\
            Before Dual-Presence & $0.167^{***}$ & $0.174^{***}$ & $-0.007$\\
            $t_{0} - 2$ & $(0.041)$ & $(0.040)$ & $(0.011)$\\
            Before Dual-Presence & $0.312^{***}$ & $0.315^{***}$ & $-0.003$ \\
            $t_{0} - 1$ & $(0.044)$ & $(0.042)$ & $(0.011)$ \\ \hline
            Start of Dual-Presence & $0.282^{***}$& $0.262^{***}$ & $0.020^{*}$ \\
            $t_{0}$ & $(0.044)$ & $(0.042)$ & $(0.011)$\\
            During Dual-Presence & $0.251^{***}$ & $0.240^{***}$ & $0.011$\\
            $t_{0} + 1$ & $(0.046)$ & $(0.044)$ & $(0.012)$\\
            During Dual-Presence & $0.268^{***}$ & $0.239^{***}$ & $0.029^{**}$\\
            $t_{0} + 2$ & $(0.047)$ & $(0.044)$ & $(0.012)$\\
            During Dual-Presence & $0.339^{***}$ & $0.315^{***}$ & $0.024^{***}$\\
            $t_{0} + 3$ to $t_{0} + 12$ & $(0.035)$ & $(0.033)$ & $(0.009)$\\ \hline
            Entry & $0.380^{***}$ & $0.356^{***}$ & $0.024^{*}$\\
            $t_{e}$ & $(0.048)$ & $(0.046)$ & $(0.012)$\\
            After Entry & $0.368^{***}$ & $0.339^{***}$ & $0.029^{**}$\\
            $t_{e} + 1$ to $t_{e} + 2$ & $(0.044)$ & $(0.042)$ & $(0.011)$\\
            After Entry & $0.240^{***}$ & $0.226^{***}$ & $0.014$\\
            $t_{e} + 3$ to $t_{e} + 12$ & $(0.045)$ & $(0.043)$ & $(0.012)$\\ \hline
            Carrier-route/time FE & Yes & Yes & Yes\\
            $R^{2}$ & $0.891$ & $0.894$ & $0.656$ \\
            $N$ & $3,157$ & $3,157$ & $3,157$\\
            \hline\hline
        \end{tabular}
        \begin{tablenotes}[flushleft]
            \item \textit{Note.} The dependent variable is the logged total passengers ($q$) or logged available seats (seats) from the T-100, or the logged load factor (the ratio of $q$ to seats). I report robust standard errors in parentheses. The asterisks $*$, $**$ and $***$ denote significance at the 90\%, 95\% and 99\% levels.
        \end{tablenotes}
    \end{threeparttable}
\end{table}

\begin{table}
    \centering
    \begin{threeparttable}
     \caption{Incumbent Response to Dual-Presence and Entry by Southwest: global network measures.}
     \label{tab:baseline_model_global_network}
\begin{tabular}{lcccccc}
            \hline\hline
            \rule{0pt}{2.5ex}
             & (1) & (2) & (3) & (4) & (5) & (6)\\
             & $\mathrm{ln}(P_{\mathrm{mean}})$ & $\mathrm{ln}(P_{\mathrm{mean}})$ & $\mathrm{ln}(P_{\mathrm{mean}})$ & $\mathrm{ln}(P_{\mathrm{mean}})$ & $\mathrm{ln}(P_{\mathrm{mean}})$ & $\mathrm{ln}(P_{\mathrm{mean}})$\\
            network measure: & density & diameter & a.p.l. & transitivity & av. clustering & assortativity\\
            \midrule
            Start of Dual-Presence & $-0.041$ & $0.107^{*}$ & $0.577^{***}$ & $-0.036$ & $-0.017$ & $-0.026$ \\
            $t_{0}$ & $(0.051)$ & $(0.057)$ & $(0.207)$ & $(0.033)$ & $(0.032)$ & $(0.070)$ \\
            $\times$ network measure & $0.117$ & $-0.049^{***}$ & $-0.300^{***}$ & $0.035$ & $0.004$ & $0.013$\\
            & $(0.558)$ & $(0.017)$ & $(0.102)$ & $(0.205)$ & $(0.057)$ & $(0.094)$ \\
            During Dual-Presence & $-0.081$ & $0.043$ & $0.345$ & $-0.069^{*}$ & $-0.058^{*}$ & $-0.010$ \\
\hline\hline
        \end{tabular}
        \begin{tablenotes}[flushleft]
            \item \textit{Note.} The dependent variable is the passenger-weighted logged mean real fare. I report robust standard errors in parentheses. The asterisks $*$, $**$ and $***$ denote significance at the 90\%, 95\% and 99\% levels.
        \end{tablenotes}
    \end{threeparttable}
\end{table}

\begin{table}
    \centering
    \begin{threeparttable}
     \caption{Incumbent Response to Dual-Presence and Entry by Southwest: local network measures.}
     \label{tab:baseline_model_local_network}
        \begin{tabular}{lcccccc}
            \hline\hline
            \rule{0pt}{2.5ex}
             & (1) & (2) & (3) & (4) & (5) & (6)\\
             & $\mathrm{ln}(P_{\mathrm{mean}})$ & $\mathrm{ln}(P_{\mathrm{mean}})$ & $\mathrm{ln}(P_{\mathrm{mean}})$ & $\mathrm{ln}(P_{\mathrm{mean}})$ & $\mathrm{ln}(P_{\mathrm{mean}})$ & $\mathrm{ln}(P_{\mathrm{mean}})$\\
            network measure: & edge between. & degree & closeness & between. & eigenv. & neighbour degree\\
            \midrule
            Start of Dual-Presence & $-0.047^{*}$ & $-0.153^{***}$ & $-0.267^{***}$ & $-0.066^{**}$ & $-0.075$ & $-0.067^{*}$ \\
            $t_{0}$ & $(0.026)$ & $(0.051)$ & $(0.085)$ & $(0.032)$ & $(0.051)$ & $(0.040)$ \\            
            $\times$ network measure & $0.476$ & $0.153^{**}$ & $0.270^{***}$ & $0.067$ & $0.119$ & $0.001$\\
            & $(0.561)$ & $(0.063)$ & $(0.100)$ & $(0.044)$ & $(0.101)$ & $(0.001)$ \\           
            During Dual-Presence & $-0.065^{**}$ & $-0.086$ & $-0.120$ & $-0.063^{*}$ & $-0.046$ & $-0.042$ \\
\hline\hline
        \end{tabular}
        \begin{tablenotes}[flushleft]
            \item \textit{Note.} The dependent variable is the passenger-weighted logged mean real fare. I report robust standard errors in parentheses. The asterisks $*$, $**$ and $***$ denote significance at the 90\%, 95\% and 99\% levels.
        \end{tablenotes}
    \end{threeparttable}
\end{table}

\begin{table}
    \centering
    \begin{threeparttable}
     \caption{Incumbent Response to Dual-Presence and Entry by Southwest: global network measures.}
     \label{tab:baseline_model_T100_global_network}
        \begin{tabular}{lcccccc}
            \hline\hline
            \rule{0pt}{2.5ex}
             & (1) & (2) & (3) & (4) & (5) & (6)\\
             & $\mathrm{ln}(q)$ & $\mathrm{ln}(q)$ & $\mathrm{ln}(q)$ & $\mathrm{ln}(q)$ & $\mathrm{ln}(q)$ & $\mathrm{ln}(q)$\\
            network measure:  & density & diameter & a.p.l. & transitivity & av. clustering & assortativity\\
            \midrule
            Start of Dual-Presence & $0.174^{*}$ & $0.521^{***}$ & $0.841^{*}$ & $0.235^{***}$ & $0.294^{***}$ & $0.125$ \\     
            $t_{0}$ & $(0.102)$ & $(0.116)$ & $(0.430)$ & $(0.064)$ & $(0.064)$ & $(0.143)$ \\
            $\times$ network measure & $0.932$ & $-0.086^{**}$ & $-0.270$ & $-0.611$ & $-0.205^{*}$ & $-0.236$\\            
            & $(1.113)$ & $(0.036)$ & $(0.211)$ & $(0.401)$ & $(0.115)$ & $(0.192)$ \\
            During Dual-Presence & $0.409^{***}$ & $0.384^{**}$ & $-0.340$ & $0.224^{***}$ & $0.283^{***}$ & $0.233$ \\
\hline\hline
        \end{tabular}
        \begin{tablenotes}[flushleft]
            \item \textit{Note.} The dependent variable is the number of passengers (T-100). I report robust standard errors in parentheses. The asterisks $*$, $**$ and $***$ denote significance at the 90\%, 95\% and 99\% levels.
        \end{tablenotes}
    \end{threeparttable}
\end{table}

\begin{table}
    \centering
    \begin{threeparttable}
     \caption{Incumbent Response to Dual-Presence and Entry by Southwest: local network measures.}
     \label{tab:baseline_model_T100_local_network}
        \begin{tabular}{lcccccc}
            \hline\hline
            \rule{0pt}{2.5ex}
             & (1) & (2) & (3) & (4) & (5) & (6)\\
             & $\mathrm{ln}(q)$ & $\mathrm{ln}(q)$ & $\mathrm{ln}(q)$ & $\mathrm{ln}(q)$ & $\mathrm{ln}(q)$ & $\mathrm{ln}(q)$\\
            network measure: & edge between. & degree & closeness & between. & eigenv. & neighbour degree\\
            \midrule
            Start of Dual-Presence & $0.198^{***}$ & $0.206^{*}$ & $0.103$ & $0.175^{***}$ & $0.137$ & $0.337^{***}$ \\     
            $t_{0}$ & $(0.052)$ & $(0.103)$ & $(0.169)$ & $(0.064)$ & $(0.098)$ & $(0.084)$ \\
            $\times$ network measure & $1.941^{*}$ & $0.010$ & $0.103$ & $0.069$ & $0.252$ & $-0.002$\\            
            & $(1.112)$ & $(0.126)$ & $(0.199)$ & $(0.089)$ & $(0.195)$ & $(0.002)$ \\
            During Dual-Presence & $0.251^{***}$ & $0.225^{*}$ & $0.122$ & $0.143^{**}$ & $0.183^{*}$ & $0.144$ \\
\hline\hline
        \end{tabular}
        \begin{tablenotes}[flushleft]
            \item \textit{Note.} The dependent variable is the number of passengers (T-100). I report robust standard errors in parentheses. The asterisks $*$, $**$ and $***$ denote significance at the 90\%, 95\% and 99\% levels.
        \end{tablenotes}
    \end{threeparttable}
\end{table}

\begin{table}
    \centering
    \begin{threeparttable}
     \caption{Incumbent Response to Dual-Presence and Entry by Southwest: global network measures.}
     \label{tab:baseline_model_T100_seats_global_network}
        \begin{tabular}{lcccccc}
            \hline\hline
            \rule{0pt}{2.5ex}
             & (1) & (2) & (3) & (4) & (5) & (6)\\
             & $\mathrm{ln}(\mathrm{seats})$ & $\mathrm{ln}(\mathrm{seats})$ & $\mathrm{ln}(\mathrm{seats})$ & $\mathrm{ln}(\mathrm{seats})$ & $\mathrm{ln}(\mathrm{seats})$ & $\mathrm{ln}(\mathrm{seats})$\\
            network measure: & density & diameter & a.p.l. & transitivity & av. clustering & assortativity\\
            \midrule
            Start of Dual-Presence & $0.208^{**}$ & $0.434^{***}$ & $0.518$ & $0.213^{***}$ & $0.261^{***}$ & $0.172$ \\     
            $t_{0}$ & $(0.097)$ & $(0.111)$ & $(0.411)$ & $(0.062)$ & $(0.061)$ & $(0.137)$ \\
            $\times$ network measure & $0.328$ & $-0.063^{*}$ & $-0.123$ & $-0.487$ & $-0.164$ & $-0.138$\\            
            & $(1.067)$ & $(0.034)$ & $(0.202)$ & $(0.387)$ & $(0.111)$ & $(0.184)$ \\
            During Dual-Presence & $0.344^{***}$ & $0.280^{*}$ & $-0.362$ & $0.184^{***}$ & $0.233^{***}$ & $0.295^{**}$ \\
\hline\hline
        \end{tabular}
        \begin{tablenotes}[flushleft]
            \item \textit{Note.} The dependent variable is the total available seats (T-100). I report robust standard errors in parentheses. The asterisks $*$, $**$ and $***$ denote significance at the 90\%, 95\% and 99\% levels.
        \end{tablenotes}
    \end{threeparttable}
\end{table}

\begin{table}
    \centering
    \begin{threeparttable}
     \caption{Incumbent Response to Dual-Presence and Entry by Southwest: local network measures.}
     \label{tab:baseline_model_T100_seats_local_network}
        \begin{tabular}{lcccccc}
            \hline\hline
            \rule{0pt}{2.5ex}
             & (1) & (2) & (3) & (4) & (5) & (6)\\
             & $\mathrm{ln}(\mathrm{seats})$ & $\mathrm{ln}(\mathrm{seats})$ & $\mathrm{ln}(\mathrm{seats})$ & $\mathrm{ln}(\mathrm{seats})$ & $\mathrm{ln}(\mathrm{seats})$ & $\mathrm{ln}(\mathrm{seats})$\\
            network measure: & edge between. & degree & closeness & between. & eigenv. & neighbour degree\\
            \midrule
            Start of Dual-Presence & $0.202^{***}$ & $0.199^{**}$ & $0.125$ & $0.173^{***}$ & $0.150$ & $0.292^{***}$ \\     
            $t_{0}$ & $(0.050)$ & $(0.098)$ & $(0.161)$ & $(0.062)$ & $(0.095)$ & $(0.080)$ \\
            $\times$ network measure & $1.210$ & $0.002$ & $0.062$ & $0.055$ & $0.193$ & $-0.001$\\            
            & $(1.067)$ & $(0.121)$ & $(0.191)$ & $(0.086)$ & $(0.189)$ & $(0.002)$ \\
            During Dual-Presence & $0.246^{***}$ & $0.241^{**}$ & $0.183$ & $0.169^{**}$ & $0.225^{**}$ & $0.158^{*}$ \\
\hline\hline
        \end{tabular}
        \begin{tablenotes}[flushleft]
            \item \textit{Note.} The dependent variable is the total available seats (T-100). I report robust standard errors in parentheses. The asterisks $*$, $**$ and $***$ denote significance at the 90\%, 95\% and 99\% levels.
        \end{tablenotes}
    \end{threeparttable}
\end{table}

\clearpage

\section{Calculation of Network Measures}\label{sec:network_illustration}

\begin{figure}[!htbp]\centering
    	\includegraphics[width=.55\linewidth, keepaspectratio]{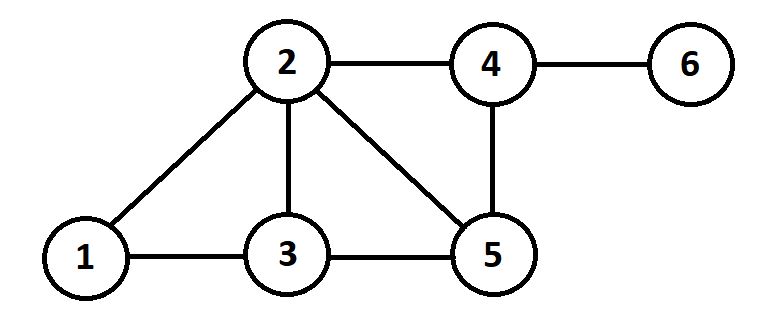}
    	\caption{An illustrative network.}
    	\label{fig:illustrative_network}
\end{figure}

\begin{table}[!htbp]
    \centering
    \begin{threeparttable}
     \caption{Global network measures.}
     \label{tab:illustrative_network_global}
        \begin{tabular}{cccccc}
            \hline\hline
            \rule{0pt}{2.5ex}
             density & diameter & a.p.l. & transitivity & av. clustering & assortativity\\
            \midrule
            $0.5\overline{3}$ & $3$ & $1.6$ & $0.5625$ & $0.52\overline{7}$ & $-0.2$\\
            \hline\hline
        \end{tabular}
    \end{threeparttable}
\end{table}

\begin{table}[!htbp]
    \centering
    \begin{threeparttable}
     \caption{Local network measures (node).}
     \label{tab:illustrative_network_local_node}
        \begin{tabular}{lcccccc}
            \hline\hline
            \rule{0pt}{2.5ex}
             & node 1 & node 2 & node 3 & node 4 & node 5 & node 6\\
            \midrule
            degree & $0.4$ & $0.8$ & $0.6$ & $0.6$ & $0.6$ & $0.2$\\
            closeness & $0.\overline{5}$ & $0.8\overline{3}$ & $0.625$ & $0.\overline{714285}$ & $0.\overline{714285}$ & $0.\overline{45}$\\
            between. & $0$ & $0.35$ & $0.05$ & $0.4$ & $0.1$ & $0$\\
            eigenv. & $0.336$ & $0.549$ & $0.453$ & $0.383$ & $0.465$ & $0.129$\\
            neighbour degree & $3.5$ & $2.75$ & $3$ & $2.\overline{6}$ & $3.\overline{3}$ & $3$\\
            \hline\hline
        \end{tabular}
        \begin{tablenotes}[flushleft]
            \item \textit{Note.} Eigenvector centrality is reported to 3 decimal places.
        \end{tablenotes}
    \end{threeparttable}
\end{table}

\begin{table}[!htbp]
    \centering
    \begin{threeparttable}
     \caption{Local network measures (edge).}
     \label{tab:illustrative_network_local_edge}
        \begin{tabular}{lcccccccc}
            \hline\hline
            \rule{0pt}{2.5ex}
            edge: & $(1,2)$ & $(1,3)$ & $(2,3)$ & $(2,4)$ & $(2,5)$ & $(3,5)$ & $(4,5)$ & $(4,6)$\\
            \midrule
            edge between. & $0.2\overline{3}$ & $0.1$ & $0.1\overline{3}$ & $0.\overline{3}$ & $0.1$ & $0.1\overline{6}$ & $0.2$ & $0.\overline{3}$\\
            degree & $0.8$ & $0.6$ & $0.8$ & $0.8$ & $0.8$ & $0.6$ & $0.6$ & $0.6$\\
            closeness & $0.8\overline{3}$ & $0.625$ & $0.8\overline{3}$ & $0.8\overline{3}$ & $0.8\overline{3}$ & $0.\overline{714285}$ & $0.\overline{714285}$ & $0.\overline{714285}$\\
            between. & $0.35$ & $0.05$ & $0.35$ & $0.4$ & $0.35$ & $0.1$ & $0.4$ & $0.4$\\
            eigenv. & $0.549$ & $0.453$ & $0.549$ & $0.549$ & $0.549$ & $0.465$ & $0.465$ & $0.383$\\
            neighbour degree & $3.5$ & $3.5$ & $3$ & $2.75$ & $3.\overline{3}$ & $3.\overline{3}$ & $3.\overline{3}$ & $3$\\
            \hline\hline
        \end{tabular}
    \end{threeparttable}
\end{table}

\end{document}

%% file: preamble.tex
\usepackage[utf8]{inputenc}

\usepackage{longtable}
\usepackage[left=2cm,right=2cm,top=3cm,bottom=2.5cm]{geometry}
\usepackage{media9} 
\usepackage{fancyhdr}
\usepackage{setspace}
\usepackage{import}
\usepackage[hang, small,labelfont=bf,up,textfont=up]{caption}
\usepackage{subcaption}

\usepackage{sectsty}
\sectionfont{\centering}
\usepackage{float}

\usepackage{blkarray}
\usepackage{graphicx}
\usepackage{mwe}

\newcommand{\motif}[2]{\vcenter{\hbox{\includegraphics[scale=0.3]{M_11_4}}}}

\usepackage{todonotes}
\usepackage[english]{babel}										
\usepackage[protrusion=true,expansion=true]{microtype}
\usepackage{gensymb}
\usepackage{booktabs}
\usepackage{mathptmx}
\usepackage{siunitx}
\usepackage{nomencl}
\usepackage{multirow}
\usepackage{framed}
\usepackage{multicol}
\usepackage{url}
\usepackage{amsmath, amsfonts,amsthm,amssymb}

\usepackage{pdflscape}

\usepackage{color}
\makeatletter
\def\BState{\State\hskip-\ALG@thistlm}
\makeatother

\usepackage{verbatim}

\usepackage{tablefootnote}

\theoremstyle{plain}

\theoremstyle{definition}

\theoremstyle{definition}

\usepackage[explicit]{titlesec}
\titleformat{name=\paragraph,numberless}[runin]
  {\normalfont\normalsize\bfseries}{}{15pt}{\uline{#1.}}
  
\usepackage[toc,page]{appendix}

\usepackage{algorithm}
\usepackage{algpseudocode}

\usepackage{footnote}
\makesavenoteenv{tabular}
\makesavenoteenv{table}

\usepackage{threeparttable}

\usepackage{siunitx}
\usepackage{dcolumn}
\newcolumntype{d}[1]{D{.}{.}{#1}}
\usepackage{etoolbox}
\robustify\bfseries
\usepackage{chngcntr}

\usepackage{array}

\usepackage{natbib}

\usepackage{enumitem}

\usepackage{mathtools}

\usepackage{sectsty}
\sectionfont{\bfseries\Large\raggedright}

\usepackage{tikz}
\newcommand\hlight[1]{\tikz[overlay, remember picture,baseline=-\the\dimexpr\fontdimen22\textfont2\relax]\node[rectangle,fill=blue!50,rounded corners,fill opacity = 0.2,draw,thick,text opacity =1] {$#1$};}

\usepackage{marginnote}

\usepackage{breqn}